\documentclass{aa}
\usepackage{graphicx}
\usepackage{xcolor}
\usepackage{orcidlink}
\usepackage[varg]{txfonts}
\usepackage{url}
\usepackage{hyperref}

\newcommand{\fcolor}{f_{\rm col}}
 
\begin{document}

\title{X-ray polarization in the soft state of Cyg X-1}
\author{A.~Nied\'zwiecki\inst{\ref{inst1}}\orcidlink{0000-0002-8541-8849}\fnmsep\thanks{andrzej.m.niedzwiecki@gmail.com}\and
M.~Szanecki\inst{\ref{inst1}}\orcidlink{0000-0001-7606-5925}\and
A.~Veledina\inst{\ref{in:UTU},\ref{in:Nordita}}\orcidlink{0000-0002-5767-7253} \and
A.~A.~Zdziarski\inst{\ref{inst3}}\orcidlink{0000-0002-0333-2452}
\and
A.~Chakraborty\inst{\ref{inst1}}\orcidlink{0009-0008-6724-2481} \and 
J.~Poutanen\inst{\ref{in:UTU}}\orcidlink{0000-0002-0983-0049} \and 
P.~Lubi\'nski\inst{\ref{in:IFUZ}}\orcidlink{0000-0001-6276-7045} \and
A. Salganik\inst{\ref{in:UTU}}\orcidlink{0000-0003-2609-8838}}

\institute{University of Lodz, Pomorska 149/153, 90-236 \L\'od\'z, Poland \label{inst1}
\and
Department of Physics and Astronomy, FI-20014 University of Turku, Finland \label{in:UTU}
\and
Nordita, KTH Royal Institute of Technology and Stockholm University, Hannes Alfv\'ens v\"ag 12, SE-10691 Stockholm, Sweden \label{in:Nordita}
\and
Nicolaus Copernicus Astronomical Center, Polish Academy of Sciences, Bartycka 18, PL-00-716 Warszawa, Poland
\label{inst3}
\and 
Institute of Physics, University of Zielona G\'ora, Szafrana 4a, PL-65-516 Zielona G\'ora, Poland
\label{in:IFUZ}
}

\date{June 2026}

  \abstract
   {} 
   {We aim to identify the physical mechanism responsible for the observed X-ray emission and polarization of Cyg X-1 in the soft state.} 
   {We performed a detailed spectral analysis of X-ray data obtained with NICER, \textit{NuSTAR}, and \textit{INTEGRAL} during observations simultaneous with \textit{IXPE} on 2023 June 20, supplemented at higher energies with archival \textit{CGRO} data. We developed a new model, {\tt retBB}, which describes thermal disk emission and its returning reflection, and applied it together with accurate models of Comptonization and relativistically broadened reflection. Using the resulting spectral solution, we computed the expected polarization signal and compared it with the \textit{IXPE} measurements.}
   {Our spectropolarimetric modeling shows that the observed polarization is in agreement with being produced by Comptonization in a corona undergoing a semi-relativistic outflow at a velocity of $\simeq 0.3c$. Our spectral solutions admit either low or high black-hole spin values, depending on the adopted model setup. However, the observed polarization strongly favors a low spin. At high spin, the polarization angle would inevitably rotate significantly across the energy band, which is inconsistent with the observations. Apart from this rotation of the polarization angle, general relativistic effects do not play a significant role in producing the observed polarization. In particular, we find that, at most, returning disk radiation contributes only a minor amount.}
   {}


\maketitle
\nolinenumbers

\section{Introduction}

Relativistic effects play a significant role in shaping the polarization of radiation from a thermal, Keplerian accretion disk orbiting a Kerr black hole \citep{1980ApJ...235..224C,1990MNRAS.242..560L,2009ApJ...701.1175S,Loktev2024}. An important aspect of these theoretical predictions is the return of disk radiation to the disk surface due to strong gravitational light bending near the black hole. If the disk surface is ionized, reflection of these returning photons produces a distinct spectral component \citep{1976ApJ...208..534C} that is strongly polarized (at the $\sim$10\% level) perpendicular to the disk plane \citep{2000ApJ...528..161A}, in contrast to the parallel polarization orientation expected for direct disk emission from an electron-scattering-dominated atmosphere. The resulting polarization properties depend sensitively on the black hole spin and have therefore been proposed as a means of diagnosing this parameter \citep[e.g.,][]{2009ApJ...701.1175S,2012ApJ...754..133K}. A key limitation of this concept, however, is that additional spectral components may mask these characteristic signatures, particularly within the 2--8 keV energy range of the \textit{Imaging X-ray Polarimetry Explorer} \citep[\textit{IXPE};][]{2022JATIS...8b6002W}.
The polarization of disk radiation is most readily probed in the soft state of black hole systems, in which the presence of a thermal disk extending close to the black hole is well established \citep{Zdziarski2004a,2007A&ARv..15....1D}. However,  in line with the above caveat, the characteristic imprints on the polarization angle (PA) expected either from relativistic effects or from the interplay between oppositely oriented polarization directions of the direct and reflected returning radiation, have not been observed in current measurements of soft-state polarization \citep[e.g.,][]{2024A&A...684A..95M,2024ApJ...964...77R,2024ApJ...969L..30S,2024ApJ...960....3S}.

In this work, we focus on Cyg X-1, which -- with its well-characterized X-ray spectrum \citep[e.g.,][]{1999MNRAS.309..496G,2024ApJ...967L...9Z}, precisely measured distance, and a black hole mass that is relatively well constrained to within a factor of $\sim$2 \citep{2021Sci...371.1046M,2025A&A...698A..37R} -- is particularly well suited for studies of accretion physics. \cite{2024ApJ...969L..30S} argue that the soft-state polarization of Cyg X-1 is dominated by returning disk and coronal radiation. If correct, this would provide strong support for a rapidly rotating black hole in this system, since the effect is negligible for slowly spinning black holes. However, attributing the observed polarization signal to this mechanism is uncertain, as even for a maximally spinning black hole, the returning radiation is expected to contribute only a small fraction of the high-energy tail observed in this source. We therefore revisit the analysis of the soft-state spectropolarimetric data of Cyg X-1. Using our new model that includes returning radiation, {\tt retBB}, we find that this effect plays only a minor role and instead explore alternative explanations for the observed polarization.

\section{The data}

For the spectral analysis, we used the same data as in epoch~5 of \cite{2024ApJ...969L..30S} from the Neutron Star Interior Composition Explorer (NICER), the \textit{Nuclear Spectroscopic Telescope Array} (\textit{NuSTAR}), and the Imager on Board the INTEGRAL  Satellite (IBIS), obtained simultaneously with one of the \textit{IXPE} observations on 2023 June 20. The IBIS consists of two detectors, ISGRI and PICSiT; here, we used the data from ISGRI.
The NICER data are affected by stray sunlight;\footnote{\url{https://heasarc.gsfc.nasa.gov/docs/nicer/}} however, the related problems seem to occur mainly above 6 keV. 
The data below 6 keV are consistent with the simultaneously acquired \textit{NuSTAR} data and with an earlier NICER observation of Cyg X-1 in the soft state.\footnote{We applied the model developed here to the simultaneous NICER/\textit{NuSTAR} observations of Cyg~X-1 obtained in 2019 and recovered the same accretion disk parameters as in the present analysis, with a similar pattern of residuals to those shown in Fig.~\ref{fig:mod2} for the current NICER data.}
Therefore, we restricted the NICER data to the 1--6 keV energy range.\footnote{We excluded the NICER data below 1 keV due to the well-known calibration uncertainties at these energies \citep[see e.g.,][and references therein]{2022MNRAS.514.6102P,2023MNRAS.526.1154M}. While restricting the NICER analysis to energies above 1 keV is, in our view, methodologically more robust, we note that \citet{2024ApJ...969L..30S} included data below 1 keV in their analysis. We therefore verified that including or excluding the 0.5--1 keV range does not affect the results obtained for the phenomenological model of \citet{2024ApJ...969L..30S} (see Sect.~\ref{sect:steiner}).}

In addition, we used data from the Oriented Scintillation Spectrometer Experiment (OSSE) on board the \textit{Compton Gamma Ray Observatory} (\textit{CGRO}). OSSE observed \mbox{Cyg X-1} twice in the soft state, once in 1994 \citep{1996ApJ...465..907P} and once in 1996 \citep{1999MNRAS.309..496G,2002ApJ...572..984M}. 
These data extend, with relatively high statistics, beyond the range available to ISGRI, up to $\approx$600~keV, and the OSSE detector is very well calibrated. 
We compared the spectral slopes of the two datasets with those of ISGRI and find that the 1994 data are very similar (except for normalization); thus, we use them here. 

The \textit{IXPE}, NICER, \textit{NuSTAR}, IBIS, and OSSE data considered for the spectral analysis are in the 2--8, 1--6, 3--78, 32--220 and 50--600~keV ranges, respectively; however, for our main results we used only the data below 100 keV (see Sect.~\ref{sect:results}). 
During the standard NICER data reduction, the script {\tt nicerl3-spect} added a 1.5\% systematic error to each channel in the 0.34--9.0 keV range and a 2.5\% systematic error beyond that. We did not add any systematic errors to the \textit{NuSTAR} data. 
The NICER and \textit{NuSTAR} spectral data are optimally binned \citep{2016A&A...587A.151K} with the additional requirement of at least 20 counts per bin. We added a 2\% systematic uncertainty to the IBIS data, as recommended by the IBIS team \citep{ibismanual} for all data taken after 2020.\footnote{Following checks against Crab Nebula spectra collected in 2023, \cite{ibismanual} recommend adding a systematic uncertainty of at least 1\%. We adopted a slightly higher value because the soft-state Cyg X-1 spectra are significantly steeper than those of the Crab Nebula and exhibit larger deviations.}

We fit the spectra using \textsc{xspec} \citep{Arnaud96} and report uncertainties at 90\% confidence. Differences between the calibrations of different detectors were taken into account by multiplying the model spectra by $KE^{-\Delta \Gamma}$ (using {\tt plabs} in \textsc{xspec}). We assumed $K = 1$ and $\Delta \Gamma = 0$ for the \textit{NuSTAR} FPMA detector, as well as $\Delta \Gamma = 0$ for ISGRI and OSSE (necessary due to the relatively large statistical uncertainties). We set the photon energy grid (as necessary for convolution models) by the \textsc{xspec} command {\tt energies 0.01 1000 1000 log}.

We compared the polarization predicted by our models with the averaged \textit{IXPE} data from seven soft-state observations of Cyg X-1 carried out in 2023 and 2024 and reported in \citet{2025A&A...701A.115K}. The primary advantage of using averaged data is the reduction of statistical uncertainties. As noted by \citet{2025A&A...701A.115K}, polarization remains relatively stable within the soft state, which supports this approach. Some variability may be associated with orbital motion, as observed in the hard state \citep[where a better coverage of orbital phases is available; see][]{2025A&A...701A.115K}, and it may also be present in the soft state, given that the underlying mechanism is not yet fully understood. Since the averaged dataset includes measurements obtained at different orbital phases, any potential orbital modulation is effectively smoothed out, making it well suited for comparison with our theoretical models.

\section{The model}

We built the spectral model closely following \cite{2024ApJ...967L...9Z}. We adopted the distance to the source, $D = 2.22$ kpc, and the black hole mass, $M_{\rm BH} = 21.2 M_{\odot}$, as determined by \cite{2021Sci...371.1046M}; however, considering a recent investigation by \cite{2025A&A...698A..37R}, who estimated a lower mass range of $M \approx $(12.7--17.8)$ M_{\odot}$, we also examined how the results depend on the assumed $M_{\rm BH}$. For the Galactic absorption, we used \texttt{tbabs}, which assumes the abundances from \citet{2000ApJ...542..914W}. 

We developed a new model, \texttt{retBB}, for the direct and returning emission from a thermal disk, and we describe and compare it with similar models in Appendix~\ref{sect:retbb}. We considered disk emission from the innermost stable circular orbit (ISCO) out to $10^3 R_{\rm g}$, where $R_{\rm g} = GM_{\rm BH}/c^2$. For the \citet{1974ApJ...191..499P} temperature profile and neglecting returning radiation, {\tt retBB} is fully consistent with the {\tt kerrbb} model \citep{2005ApJS..157..335L}. The models differ in how they treat returning photons: in {\tt kerrbb} they are assumed to be completely absorbed and thereby increase the local temperature of the disk, whereas their reflection is accounted for in {\tt retBB}. 
The reflection of blackbody photons was computed using the {\tt xillverNS} tables developed for that case by \citet{2022ApJ...926...13G}. 
To enable comparison with other models, {\tt retBB} also enables fully elastic reflection. This is computed using the formalism of \citet{1960ratr.book.....C}. 
In the \textsc{xspec} implementation of either version of the model, we only considered first-order returning photons. The contribution from higher orders is small (see Appendix~\ref{sect:retbb} for details), and including it would make the model very slow. Moreover, no publicly available opacity tables are suitable for higher reflection orders. {\tt retBB} is parameterized by the black hole mass ($M_{\rm BH}$) and spin ($a$), accretion rate ($\dot M$), inclination angle ($i$), and the spectral hardening factor ($\fcolor = T_{\rm col}/T_{\rm eff}$), where $T_{\rm col}$  is the local color temperature and $T_{\rm eff}= (\mathcal{F}/{\sigma_{\rm SB}})^{1/4}$ is the effective temperature. Here, $\mathcal{F}$ is the local radiation flux and $\sigma_{\rm SB}$ is the Stefan-Boltzmann constant. For the {\tt xillverNS} reflection, the disk surface is characterized by the ionization parameter ($\xi$), iron abundance ($Z_{\rm Fe}$), and density ($n$). 
The last parameter was allowed to vary in the range $10^{16}$--$10^{18}\,\mathrm{cm^{-3}}$. Although {\tt xillverNS} covers a broader range of densities, we verified that the model fails to conserve energy (in the sense that the angle-averaged reflected energy exceeds the incident energy\footnote{We find that in some cases, for $n = 10^{15}$ or $10^{19} \, \mathrm{cm^{-3}}$, the angle-averaged reflected flux, integrated over the full energy range of the model, exceeds the incident flux by more than two orders of magnitude.}) for $n < 10^{16}\,\mathrm{cm^{-3}}$ and $n > 10^{18} \, \mathrm{cm^{-3}}$.

\begin{figure}
\resizebox{\hsize}{!}{
\begin{tabular}{@{}cc@{}}
\includegraphics[height=4.4cm]{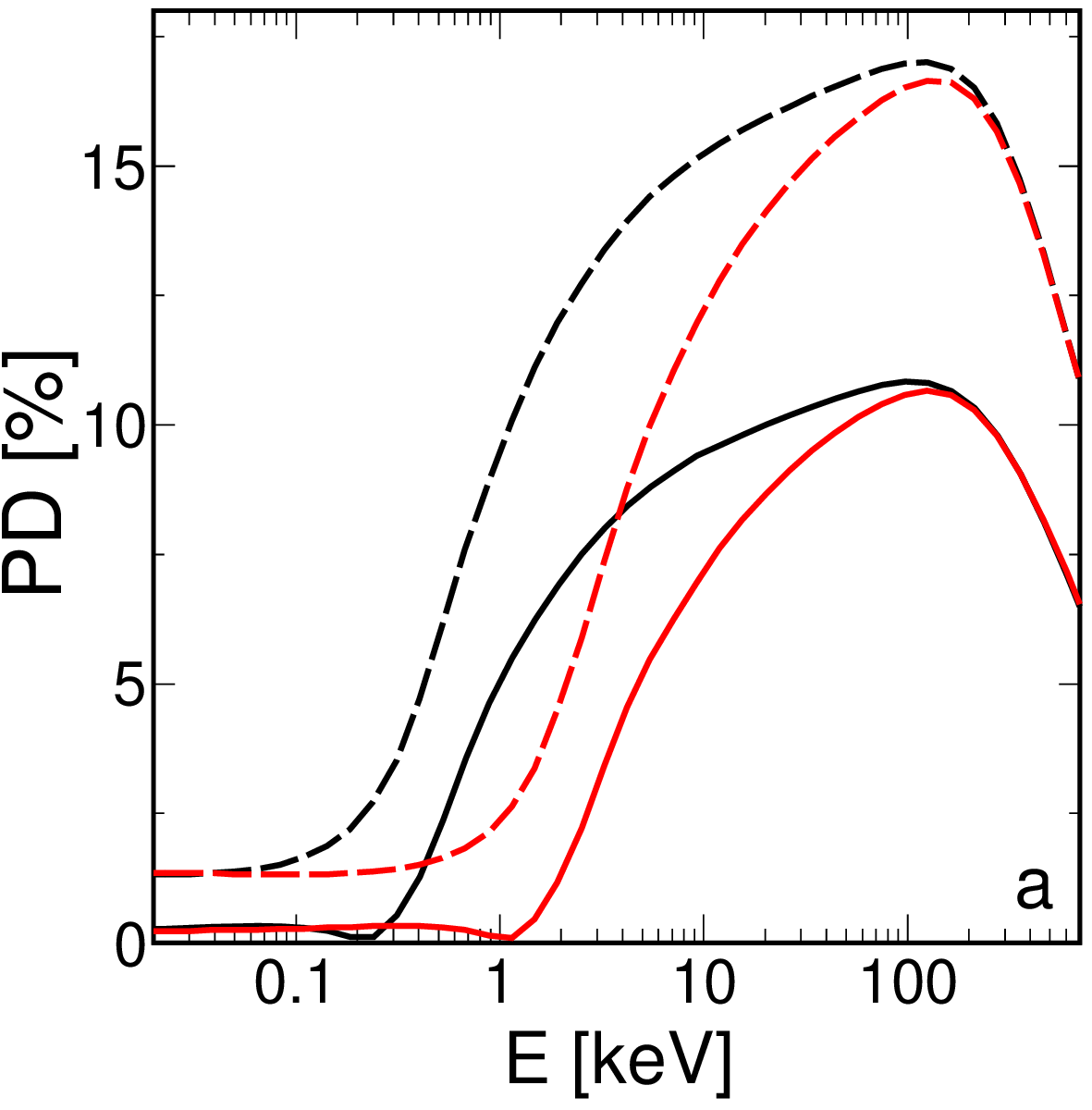}
&
\includegraphics[height=4.4cm]{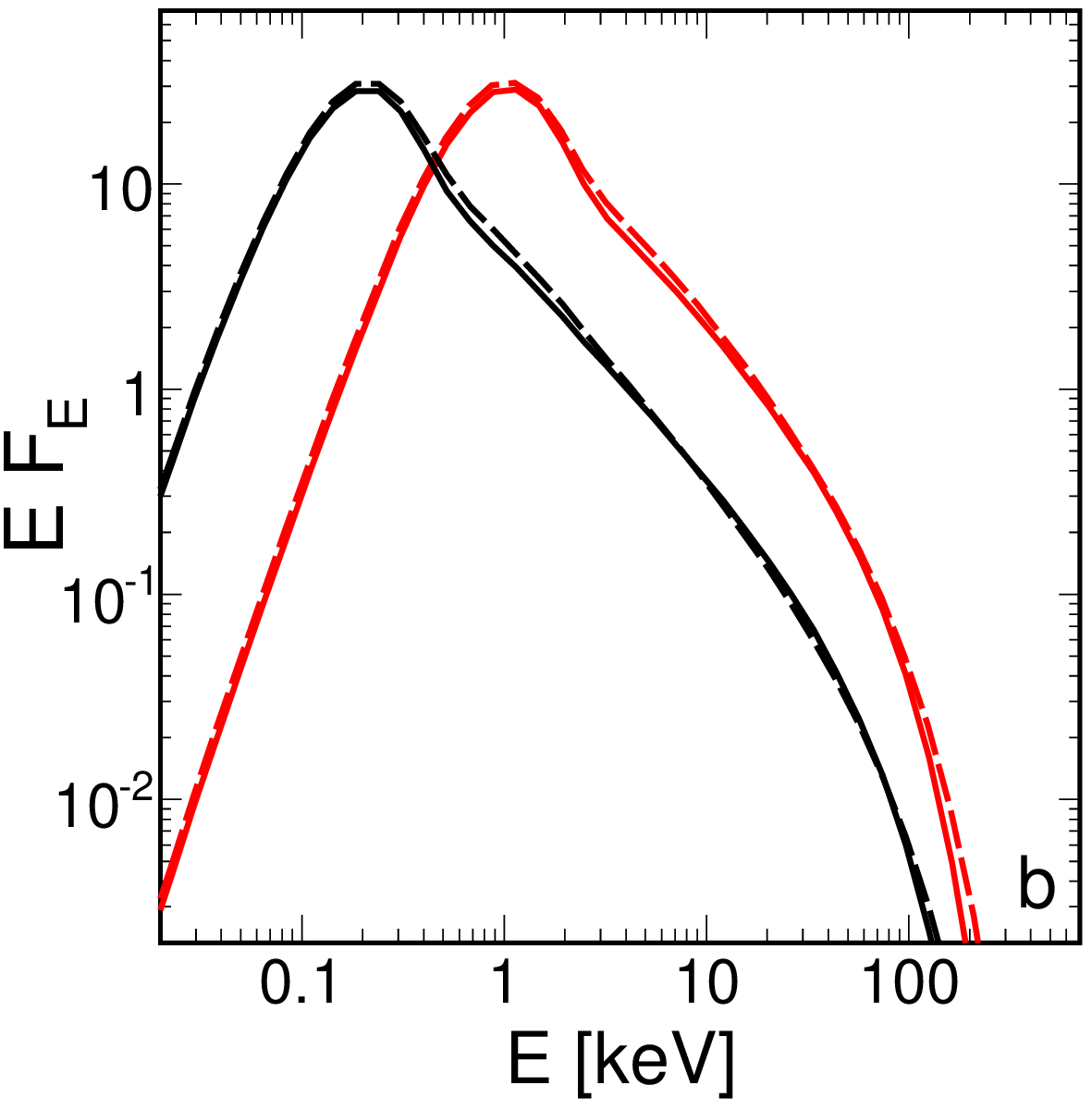} \\
\includegraphics[height=4.4cm]{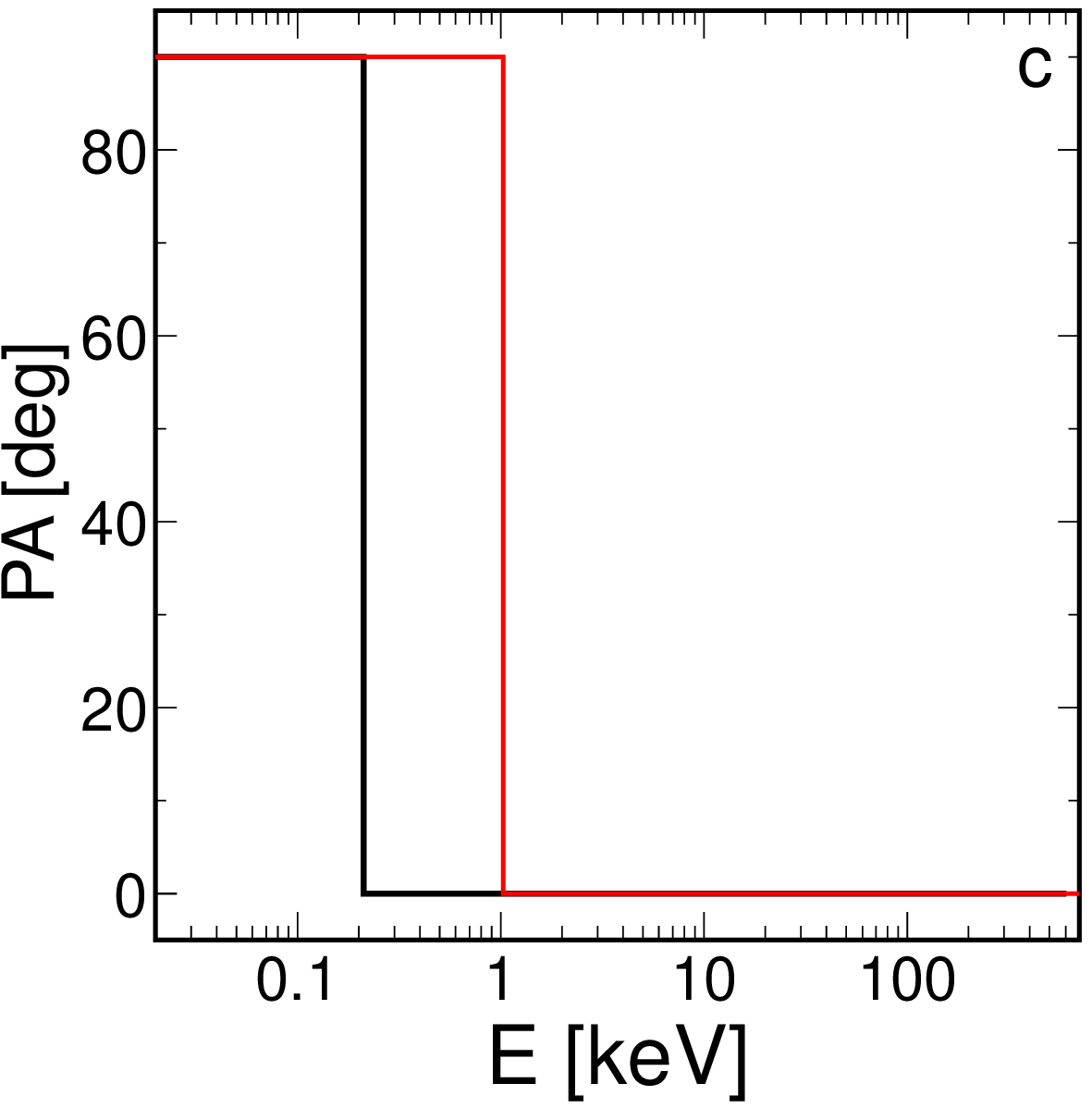} &
\end{tabular}
}
\caption{Spectropolarimetric properties of radiation, Comptonized in a slab geometry representing the disk-corona system. (a) Energy dependence of the PD as observed at an inclination $i = 35\degr$ for the outflow velocities $v = 0$ (solid) and $v=0.3c$ (dashed) and two values of the blackbody temperature of seed photons, $kT_{\rm seed} = 0.05$~keV (black) and 0.25~keV (red), corresponding to the color temperatures in our fitted disk model at $\simeq 100 R_{\rm g}$ and $\simeq 10 R_{\rm g}$, respectively. 
In all cases, the Comptonization parameters are $kT_{\rm e} = 25$~keV and $\tau = 0.75$, which correspond to the corona parameters found in Sect.~\ref{sect:results}. 
(b) Corresponding energy spectra. (c) Corresponding PA for $v=0$; for $v=0.3c$, $\mathrm{PA} = 0\degr$ at all energies.}
\label{fig:outflow}
\end{figure}

\begin{table}
\caption{Model definitions.}
\centering
\begin{tabular}{l}
\hline
\noalign{\smallskip}
Model 1: hot corona \\ 
\noalign{\smallskip}
{\tt DE}: {\tt comppsc}$\cdot${\tt retBB(elastic)} \\ 
\noalign{\smallskip} 
\hline 
\noalign{\smallskip}
Model 2: hot corona \\ 
\noalign{\smallskip}
{\tt DE}: {\tt comppsc}$\cdot${\tt retBB(xillverNS)} \\ 
\noalign{\smallskip} 
\hline 
\noalign{\smallskip}
Model 3: hot and warm corona \\ 
\noalign{\smallskip}
{\tt DE}: {\tt comppsc}$\cdot${\tt thcomp}$\cdot${\tt retBB(xillverNS)} \\ 
\noalign{\smallskip} 
\hline
\noalign{\smallskip}
Model 4: no corona \\ 
\noalign{\smallskip}
{\tt DE}: {\tt retBB(xillverNS)} \\ 
\noalign{\smallskip}
\hline
\noalign{\smallskip}
Model 5: hot corona, reflection-dominated ($\mathcal{R} \ge 7$) \\ 
\noalign{\smallskip}
{\tt DE}: {\tt comppsc}$\cdot${\tt retBB(xillverNS)} \\ 
\noalign{\smallskip}
\hline
\end{tabular}
\tablefoot{The models are constructed as {\tt plabs}$\cdot${\tt tbabs}$\cdot$({\tt DE} + {\tt RR} + {\tt SR}), where {\tt DE} denotes the direct emission (i.e., the sum of hot-corona Comptonization and unscattered thermal disk emission, including reflected returning radiation); {\tt RR} is the relativistic reflection component; and {\tt SR} is the static reflection component. In all models, {\tt SR} is computed as {\tt xilconv}$\cdot${\tt DE} and {\tt RR} is computed as {\tt relconv}$\cdot${\tt comppsc}$\cdot${\tt xilconv}$\cdot${\tt DE}, except for model~4, in which the relativistic reflection component is not included. As indicated in the {\tt DE} terms, reflection of returning radiation is treated as elastic in model 1 and with {\tt xillverNS} in models 2--5. {\tt comppsc} represents the hot corona Comptonization, affecting both the disk emission and the disk reflection. In model~3, {\tt thcomp} represents warm corona Comptonization.}
\label{tab:models_def}
\end{table}

Following \citet{2024ApJ...967L...9Z}, we considered possible dissipation in the surface layers of the disk. 
We modeled this by fully covering the disk with a warm scattering corona with $kT_{\rm e} \sim 1$ keV and $\tau \gg 1$. 
To do so, we used the {\tt thcomp} code \citep{Zdziarski2020} with free $kT_{\rm e}$ and $\tau$.

Cyg X-1 in the soft state shows relatively strong high-energy tails \citep[e.g.,][]{1999MNRAS.309..496G,2002ApJ...572..984M}. 
They are most likely produced by Comptonization by electrons in a hot corona above the disk. We assumed the seed photons originate from the disk emission. We used a convolution Comptonization code, which allows for a hybrid electron distribution, i.e., Maxwellian with a high-energy tail, {\tt comppsc}.\footnote{https://github.com/mitsza/compps\_conv} This code is a convolution version of the {\tt compps} iterative-scattering code \citep{PS96}. 
The present model approximates the high-energy tail as a power law in the electron momentum, parameterized by its index ($p$) and the minimum and maximum Lorentz factors, $\gamma_{\rm min}$ and $\gamma_{\rm max}$, respectively. 
The Maxwellian and the power law intersect at $\gamma_{\rm min}$. 
The other parameters are $kT_{\rm e}$ and $\tau$. 
We assumed a slab geometry, and we describe the covering fraction of the disk surface by the corona by $f_{\rm sc}$. 

\begin{table*}
\caption{Fitted parameter values.} 
\centering
\begin{tabular}{lllllll}
\hline \noalign{\smallskip}
\noalign{\smallskip}
 &  & model 1  & model 2 & model 3 & model 4 & model 5\\
 \noalign{\smallskip}
\hline \noalign{\smallskip}
{\tt tbabs} & $N_{\rm H}$ $[10^{21}$\,cm$^{-2}]$ & $7.6^{+0.1}_{-0.1}$   & $7.6^{+0.1}_{-0.1}$    & $8.5^{+0.1}_{-0.1}$ & $6.4^{+0.2}_{-0.2}$ & $7.0^{+0.1}_{-0.1}$ \\
\noalign{\smallskip}
\hline  
\noalign{\smallskip}
disk  & $\dot{M}\; [10^{18}\;{\rm g}\, {\rm s}^{-1}]$ & $0.20^{+0.01}_{-0.01}$ & $0.20^{+0.03}_{-0.01}$ & $0.49^{+0.04}_{-0.05}$ & $0.24^{+0.03}_{-0.04}$ & $0.21^{+0.01}_{-0.01}$\\
\noalign{\smallskip}
 & $a$                             & $0.99^{+0.01}_{-0.03}$ & $0.99^{+0.01}_{-0.01}$ & $0^{+0.2}_{-0}$   & $0.998^{+0}_{-0.001}$ & $0.98^{+0.01}_{-0.03}$ \\
 \noalign{\smallskip}
 & $i$ [deg]                  & $35^{+1}_{-1}$        & $35^{+1}_{-1}$        & $35^{+1}_{-1}$    & $35^{+3}_{-5}$ & $31^{+3}_{-2}$ \\
 \noalign{\smallskip}
      & $\fcolor$                                    & $1.5^{+0.1}_{-0.1}$    & $1.5^{+0.1}_{-0.1}$   & $1.5^{+0.1}_{-0.3}$  & $1.7^{+0.1}_{-0.1}$ & $1.7^{+0.2}_{-0.1}$ \\
      \noalign{\smallskip}
      & $Z_{\rm Fe}$                                     & --                  & $5.7^{+0.3}_{-0.6}$   & $6.0^{+0.3}_{-0.7}$ & $10^{+0}_{-2}$ & $2.8^{+0.4}_{-0.4}$ \\
      \noalign{\smallskip}
      & $\log_{10}\xi$                                  & --                  & $4.0^{+0}_{-0.1}$   & $4.0^{+0}_{-0.1}$ & $4^{+0}_{-1}$ & $4^{+0}_{-0}$ \\
      \noalign{\smallskip}
      & $\log_{10}n$                          & --                  & $18^{+0}_{-2}$ & $18^{+0}_{-2}$ & $18^{+0}_{-1}$ & $16^{+2}_{-0}$ \\
      \noalign{\smallskip}
\noalign{\smallskip}      
\hline
\noalign{\smallskip}
hot corona  & $kT_{\rm e}$\,[keV]   & $25^{+4}_{-2}$        & $25^{+2}_{-1}$         & $24^{+2}_{-1}$  & -- & $9.1^{+0.5}_{-0.3}$\\
\noalign{\smallskip}
                & $\tau$                  & $0.77^{+0.05}_{-0.05}$ & $0.77^{+0.05}_{-0.05}$  & $0.75^{+0.11}_{-0.05}$  & -- & $2.8^{+0.1}_{-0.3}$ \\
                \noalign{\smallskip}
                & $p$                  & $1.0^{+0.5}_{-0}$     & $1.0^{+0.1}_{-0}$      & $1.1^{+0.2}_{-0.1}$  & -- & $1.0^{+0.5}_{-0}$ \\
                \noalign{\smallskip}
                & $\gamma_{\rm min}$     & $1.4^{+0.1}_{-0.1}$   & $1.4^{+0.1}_{-0.1}$     & $1.5^{+0.2}_{-0.1}$ & -- & $1.2^{+0.2}_{-0.1}$\\
                \noalign{\smallskip}
                & $\gamma_{\rm max}$     & $5.1^{+0.2}_{-0.1}$   & $5.1^{+0.1}_{-0.1}$     & $5.5^{+0.1}_{-0.1}$ & -- & $17^{+1}_{-1}$\\
                \noalign{\smallskip}
                &$f_{\rm sc}$           & $0.43^{+0.04}_{-0.04}$ & $0.43^{+0.03}_{-0.05}$  & $0.41^{+0.03}_{-0.05}$ & -- & $0.013^{+0.003}_{-0.004}$\\
                \noalign{\smallskip}
\hline
\noalign{\smallskip}
warm corona    & $\tau$               & --  & -- & $23^{+4}_{-3}$  & -- & -- \\
\noalign{\smallskip}
               & $kT_{\rm e}$\,[keV]   & --  & -- & $0.49^{+0.01}_{-0.02}$  & -- & -- \\
               \noalign{\smallskip}
\hline
\noalign{\smallskip}
relat. refl.   &${\cal R}$     & $0.34^{+0.03}_{-0.02}$ & $0.34^{+0.04}_{-0.02}$ & $0.31^{+0.02}_{-0.02}$ & -- & $7.0^{+0.5}_{-0}$\\
\noalign{\smallskip}
 & $a$                             & $0.99^{+0.01}_{-0.03}$ (l)& $0.99^{+0.01}_{-0.01}$ (l)& $0^{+0.2}_{-0}$ (l)  & -- & $0.98^{+0.01}_{-0.03}$ (l)\\
 \noalign{\smallskip}
 & $i$ [deg]                    & $35^{+1}_{-1}$   (l)     & $35^{+1}_{-1}$     (l)   & $35^{+1}_{-1}$  (l)  & -- & $31^{+3}_{-2}$ (l) \\
 \noalign{\smallskip}
               & $\beta$           & $2.8^{+0.1}_{-0.1}$    & $2.8^{+0.1}_{-0.1}$   & $4.0^{+0.2}_{-0.3}$ & -- & $2.0^{+0.1}_{-0.2}$ \\
               \noalign{\smallskip}
               & $Z_{\rm Fe}$    & $5.7^{+0.3}_{-0.7}$    & $5.7^{+0.3}_{-0.6}$ (l) & $6.0^{+0.3}_{-0.7}$ (l) & -- & $2.8^{+0.4}_{-0.4}$ (l) \\
               \noalign{\smallskip}
               &$\log_{10}\xi$  & $4.0^{+0.1}_{-0.2}$    & $4.0^{+0.1}_{-0.1}$ (l) & $4.1^{+0.1}_{-0.2}$ (l) & -- & $4.37^{+0.02}_{-0.02}$ (l)\\
               \noalign{\smallskip}
\hline
\noalign{\smallskip}
static refl. & ${\cal R}$ & $0.05^{+0.01}_{-0.01}$ & $0.05^{+0.01}_{-0.01}$ & $0.04^{+0.01}_{-0.01}$ & $<1$ & $0.7^{+0.1}_{-0.1}$ \\
\noalign{\smallskip}
\hline
\noalign{\smallskip}
& $\chi_\nu^2$  & 665/637 & 660/636 & 654/634 & 536902/170 & 761/636\\
\noalign{\smallskip}
\hline
\end{tabular}
\tablefoot{Models 1, 2, 3, and 5 are fitted to the {NICER}, \textit{NuSTAR}, {ISGRI}, and {OSSE} data. Model 4 is fitted to {NICER} and \textit{NuSTAR} data only. 
In all models $M_{\rm BH} = 21.2 M_{\odot}$ and  $D = 2.22$\,kpc. (l) denotes a linked parameter; the $\xi$ parameters listed under the columns labeled "disk" and "relativistic reflection" differ due to limitations of \texttt{xillverNS}; see text. 
}
\label{tab:models}
\end{table*}

To compute the reflected spectrum, we used the convolution model, \texttt{xilconv}, developed by C. Done following the methodology of \cite{2011MNRAS.416..311K}. It is based on the \texttt{xillver} opacity tables of \cite{2013ApJ...768..146G} and the Green’s functions of \cite{1995MNRAS.273..837M}. 
The $\xi$ and $Z_{\rm Fe}$ parameters from {\tt xilconv} and {\tt retBB} are linked in our models. However, {\tt xilconv} allows the ionization parameter to vary in the $0 \le \log_{10} \xi_{\rm xil} \le 4.7$ range,  whereas {\tt xillverNS} allows the ionization parameter to vary in the $0 \le \log_{10} \xi_{\rm ret} \le 4$ range. Hence, we used $\log_{10} \xi_{\rm ret} = \min (4, \log_{10} \xi_{\rm xil})$.
We also note that {\tt xilconv} assumes $n=10^{15}$ cm$^{-3}$, which is beyond the range of densities permitted in {\tt retBB}. The reflection features are relativistically broadened, which we treated using the convolution model {\tt relconv} \citep{2010MNRAS.409.1534D}. This model assumes a power-law radial distribution of the disk irradiation, $\propto R^{-\beta}$; thus, we allowed $\beta$ to be a free parameter. Although the coronal emission approximately follows the radial dependence of the disk through the seed-photon input, the resulting reflection can deviate from this dependence because of corona-to-disk transfer effects. These may include relativistic energy shifts and light bending, particularly the return of coronal radiation initially emitted away from the disk. Such effects were not explicitly modeled here and were therefore effectively absorbed into the fitted value of $\beta$. The other {\tt relconv}  parameters, $a$ and $i$, are linked to {\tt retBB}. We assumed the inner radius of the reflection emission to be at the ISCO. We accounted for scattering of the reflected photons in the hot corona. The normalization of the reflected spectrum relative to the incident spectrum is given by the scaling parameter, $\mathcal{R}$, which for {\tt xilconv} is defined in the same way as in {\tt ireflect}.

\begin{figure}
\resizebox{\hsize}{!}{\includegraphics[width=0.95\columnwidth]{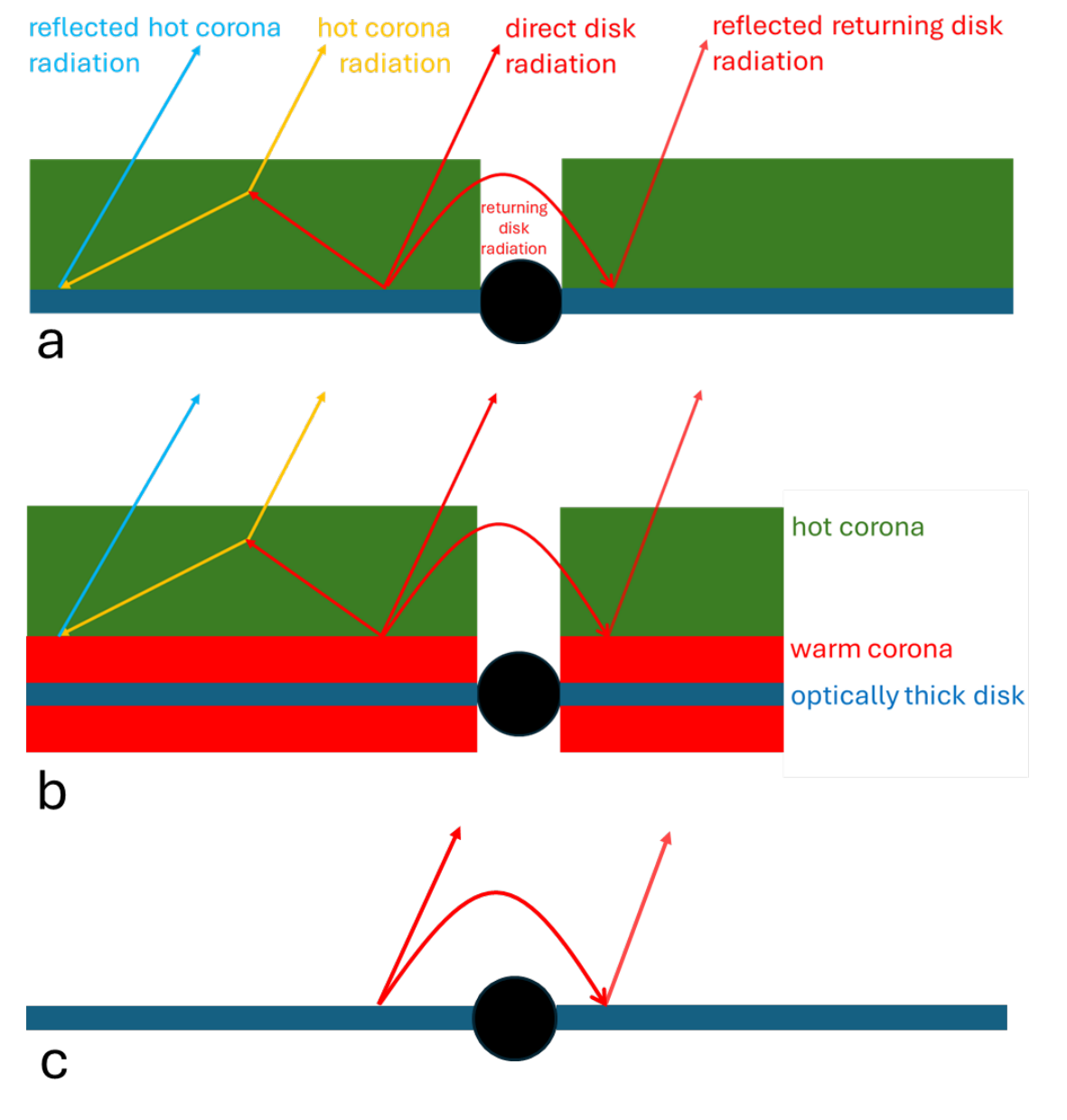}}
\caption{Geometries considered in this work and the main spectral components produced close to the black hole. (a) Geometry considered in models 1 and 2. (b) Geometry with an additional warm corona considered in model 3. (c) "No-corona" geometry considered in model 4. In panels (a) and (b), the hot corona is shown as a uniform medium fully covering the disk surface for illustrative simplicity. However, the covering factor inferred from our fits is $f_{\rm sc} \lesssim 0.5$, implying that the actual geometry is more likely characterized by localized active regions covering $\lesssim 50$\% of the disk surface.}
\label{fig:geometry}
\end{figure}

\begin{figure}
\resizebox{\hsize}{!}{
\includegraphics[width=0.85\columnwidth]{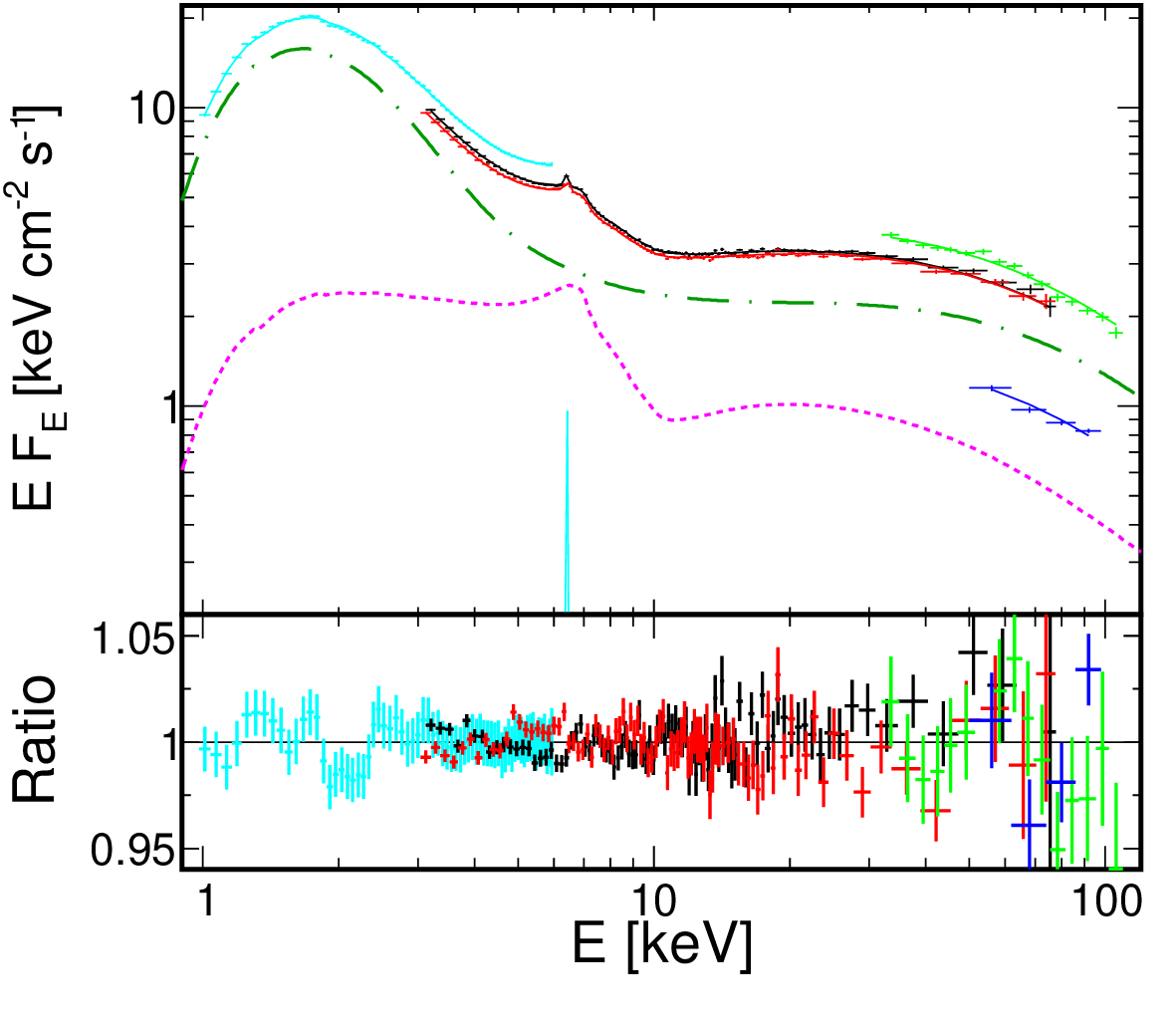}}
\caption{Spectral energy distribution of Cyg X-1. 
Top: {NICER} (cyan), \textit{NuSTAR} (black and red), \textit{INTEGRAL} (green), and \textit{CGRO} (blue) unfolded spectra for model 2. The dot-dashed (green), dotted (magenta), and solid (cyan) curves show the absorbed Comptonized component (including unscattered seed photons from the thermal disk), coronal reflection from accretion disk, and static reflection, respectively; see Table \ref{tab:models} for the definition of the spectral components. The spectral components are normalized to {NICER}. Bottom: Data-to-model ratios. }
\label{fig:mod2}
\end{figure}

Cyg X-1 also exhibits a weak, narrow Fe K$\alpha$ line, with an equivalent width of $\simeq $10~eV, which may be produced through reflection from the surface of the donor \citep{2016ApJ...826...87W}. We described it by setting $\xi_{\rm xil}=1$ in {\tt xilconv}, and we refer to this component as the static reflection.

We characterized polarization using the polarization degree (PD) and angle, ${\rm PD} = (q^2+u^2)^{1/2}$ and ${\rm PA} = 0.5{\rm atan2}(u/q)$, where $q= Q/I$ and $u = U/I$ are the normalized Stokes parameters \citep{1960ratr.book.....C}. We adopted the IAU definition, with the PA measured counterclockwise in the image plane from the projection of the black-hole spin axis. In this convention, polarization parallel to the disk plane corresponds to ${\rm PA} = \pm 90\degr$. 
Furthermore, we adopted a clockwise disk rotation for consistency with Cyg X-1, whose system exhibits clockwise orbital motion \citep{2021Sci...371.1046M}, whereas most theoretical studies assume a counterclockwise disk rotation with respect to the observer. However, following the standard practice in X-ray studies of Cyg X-1, we provide inclination angles $i<90\degr$. In the standard binary-system convention, in which clockwise orbits have inclinations $>90\degr$, this corresponds to an orbital inclination of $180\degr - i$.

\begin{figure*}
\resizebox{\hsize}{!}{\includegraphics[height=4.8cm]{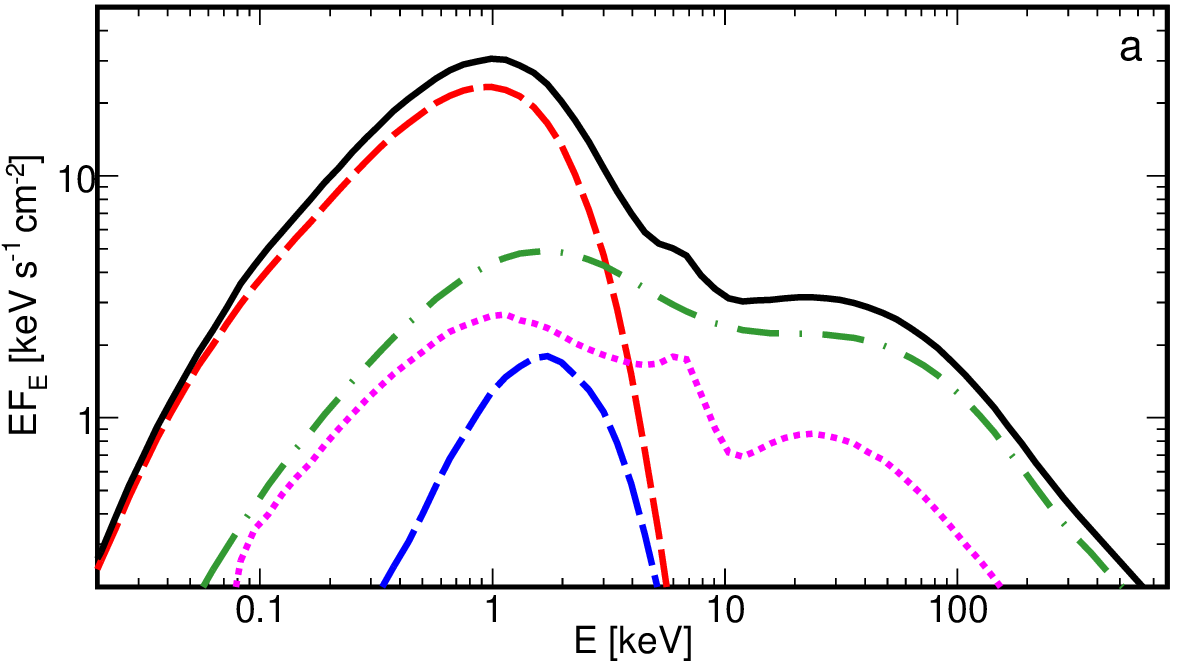} 
\includegraphics[height=4.8cm]{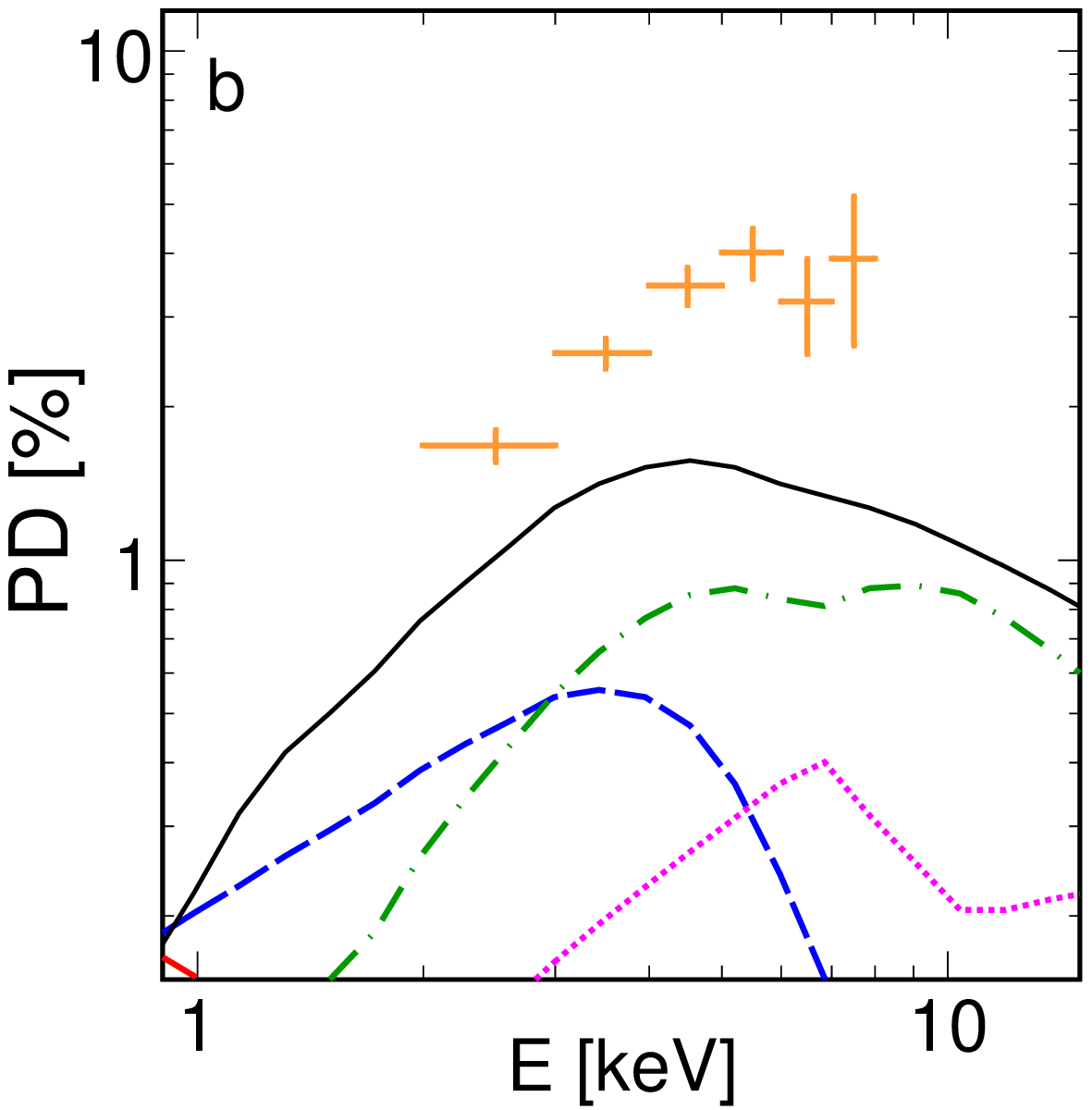}\hspace{0.1cm} \includegraphics[height=4.8cm]{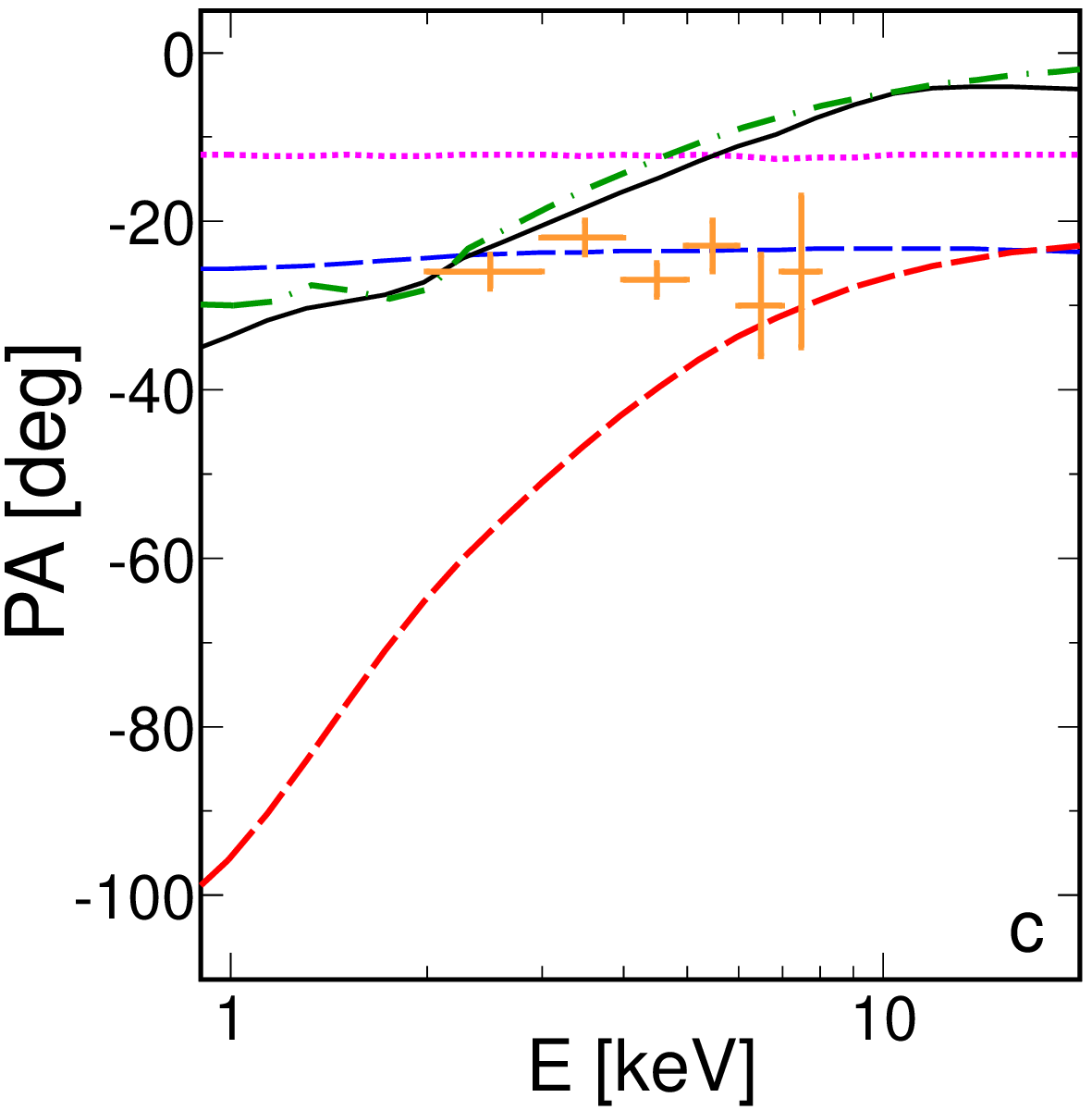}}
\caption{Unabsorbed spectral components and polarization properties of model 2 involving a rapidly rotating black hole. Polarization is calculated neglecting an outflow (i.e., $v=0$). The orange points in panels (b) and (c) show the average PD and PA measured by \textit{IXPE} in the soft state of Cyg X-1 \citep{2025A&A...701A.115K}.
(a) Total spectrum (solid black curve) and individual spectral components of model 2. The direct thermal disk emission, reflected returning radiation, hybrid Comptonized emission, and its reflection are shown by the dashed red, dashed blue, dot-dashed green, and dotted magenta curves, respectively.
(b) PD. The solid black curve shows the net PD resulting from the combined polarization signals of all components, while the colored curves show the contributions from the polarized part of each component to the total flux.
(c) PA, measured counterclockwise from celestial north. For the PA calculation, we assume that the disk rotates clockwise and that the disk axis is aligned with the radio jet, adopting a position angle of $-23\fdg5$. The line styles and colors in panels (b) and (c) correspond to those used in panel (a).
}
\label{fig:components2}
\end{figure*}

We assumed that the polarization of the direct disk emission is described by the results of a plane-parallel, electron-scattering-dominated atmosphere \citep{1960ratr.book.....C,1963trt..book.....S}. 
The same assumption was adopted for the model that includes a warm corona, in which the direct component was computed as {\tt thcomp}$\cdot${\tt retBB}. The polarization of the reflected components was calculated using the formalism of \cite{1960ratr.book.....C} for reflection from an electron-scattering-dominated slab. 
For the reflection of hot-corona radiation, modeled as {\tt xilconv}$\cdot${\tt comppsc}, we assumed locally isotropic irradiation of the disk surface. In contrast, for the returning-radiation component in {\tt retBB} we accounted for the full angular dependence of the incident flux.

The polarization of the hot-corona emission was computed using a newly developed version of {\tt compps} \citep{PS96,2023ApJ...949L..10P}, which treats the polarization of Comptonized radiation produced in a slab geometry irradiated from below. 
The model version employed here assumes only a thermal electron distribution; consequently, polarization arising from the nonthermal component of the fitted hybrid electron distribution is neglected. 
In calculating the polarization of this component, we assumed that the corona corotates with the disk and that the radius-dependent seed-photon input is described by a blackbody spectrum with the disk's local color temperature. 
We also accounted for motion in the corona away from the disk, which was previously found to be necessary to reproduce the observed hard-state polarization \citep{2023ApJ...949L..10P} and also appears necessary to account for the observed soft-state polarization within our best-fitting spectral model.
The outflow velocity measured in the corona frame is denoted by~$v$. 
Figure~\ref{fig:outflow} illustrates the dependence of PD and PA on the seed-photon temperature and the outflow velocity. 
For $v = 0$, first-order scattering produces radiation weakly polarized parallel to the slab, while higher-order scattering yields perpendicular polarization as PD increases with energy. 
For $v = 0.3c$, the radiation is polarized perpendicular to the slab at all energies, and the PD is higher than in the static case. 
A detailed discussion of the underlying physical mechanisms can be found in \cite{2023ApJ...949L..10P}. 
In our modeling, the spectral solution was obtained assuming a static corona, and the outflow was included only in the polarization calculations. 
This is justified for the semi-relativistic velocity considered here, as illustrated in Fig.~\ref{fig:outflow}. 
While $v=0.3c$ significantly increases the PD, the corresponding change in the total radiation spectrum is negligible.
 
The procedures described above yield local polarization of the reflected, Comptonized, and direct disk radiation. The transfer functions defined in Appendix~\ref{sect:retbb} were then applied to compute the polarized fluxes of those components received by a distant observer.

\section{Results}

\label{sect:results}

\begin{figure*}
\resizebox{\hsize}{!}{\includegraphics[height=4.6cm]{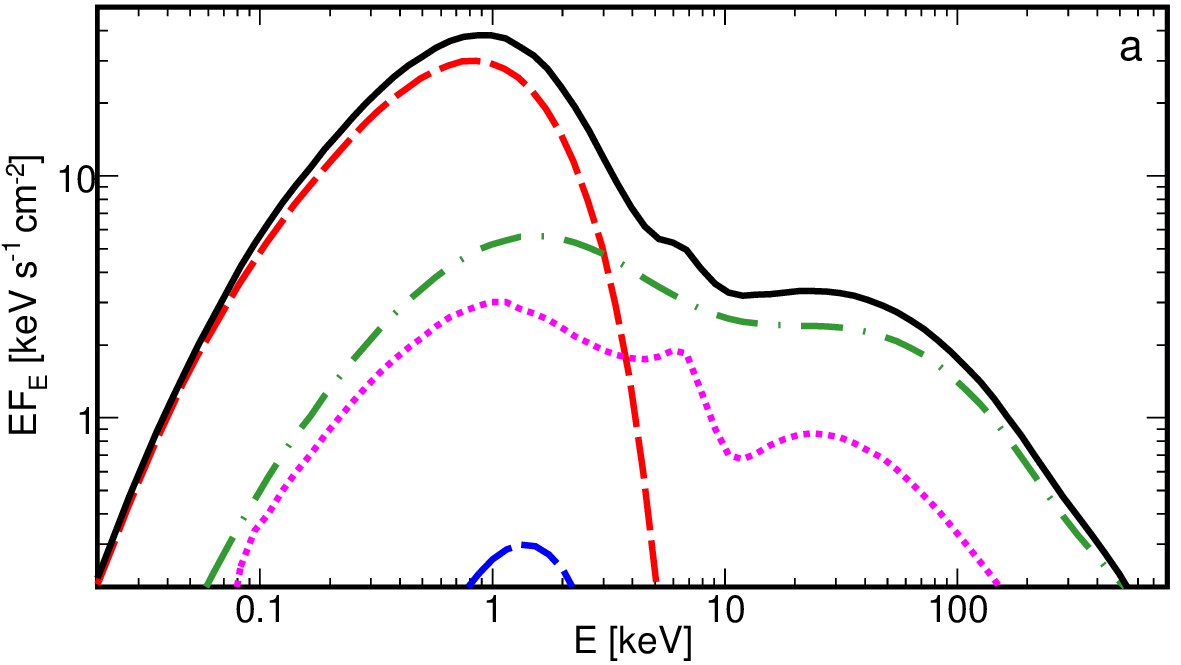} \hspace{0.1cm}\includegraphics[height=4.6cm]{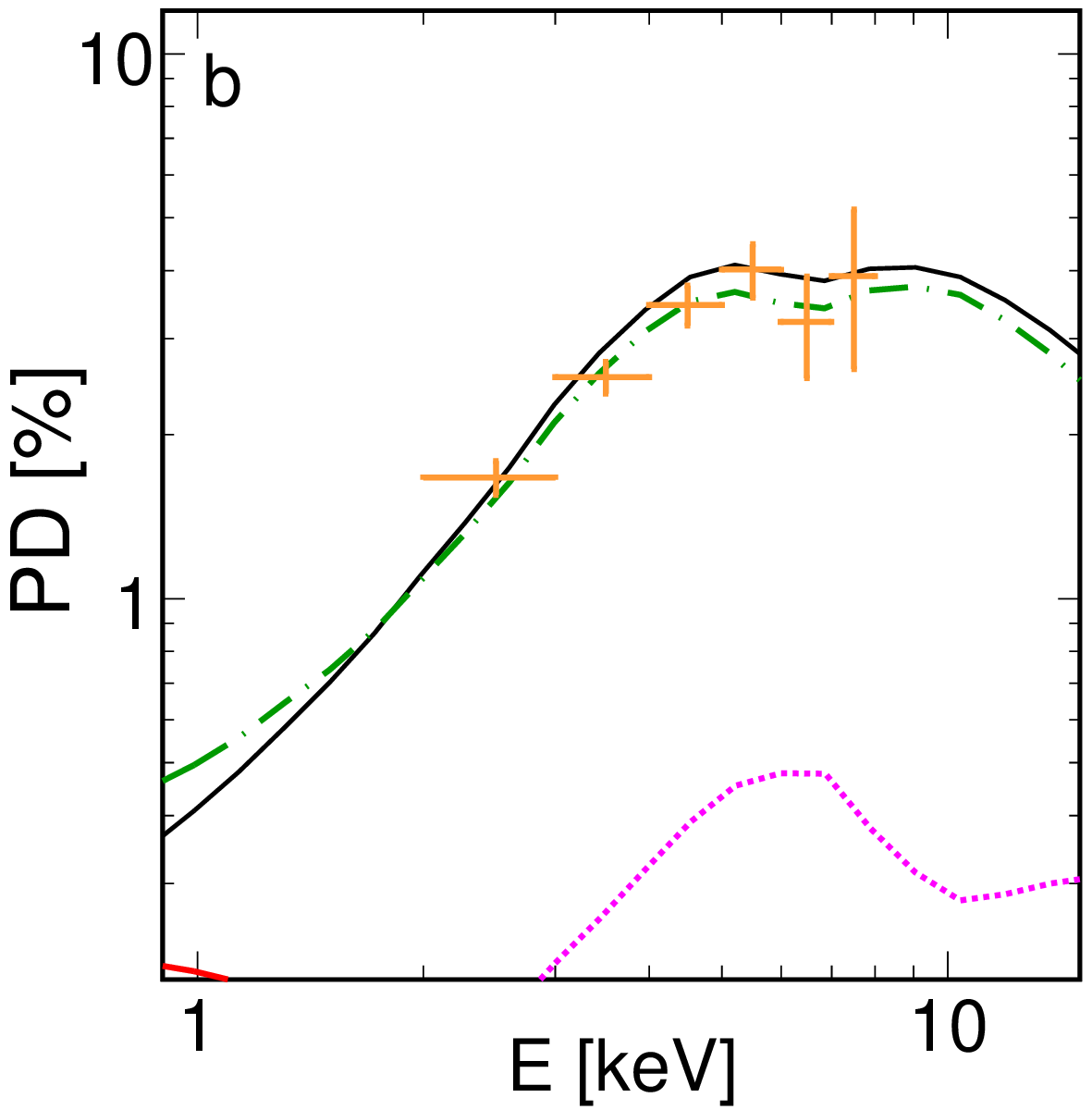}
\hspace{0.1cm}\includegraphics[height=4.6cm]{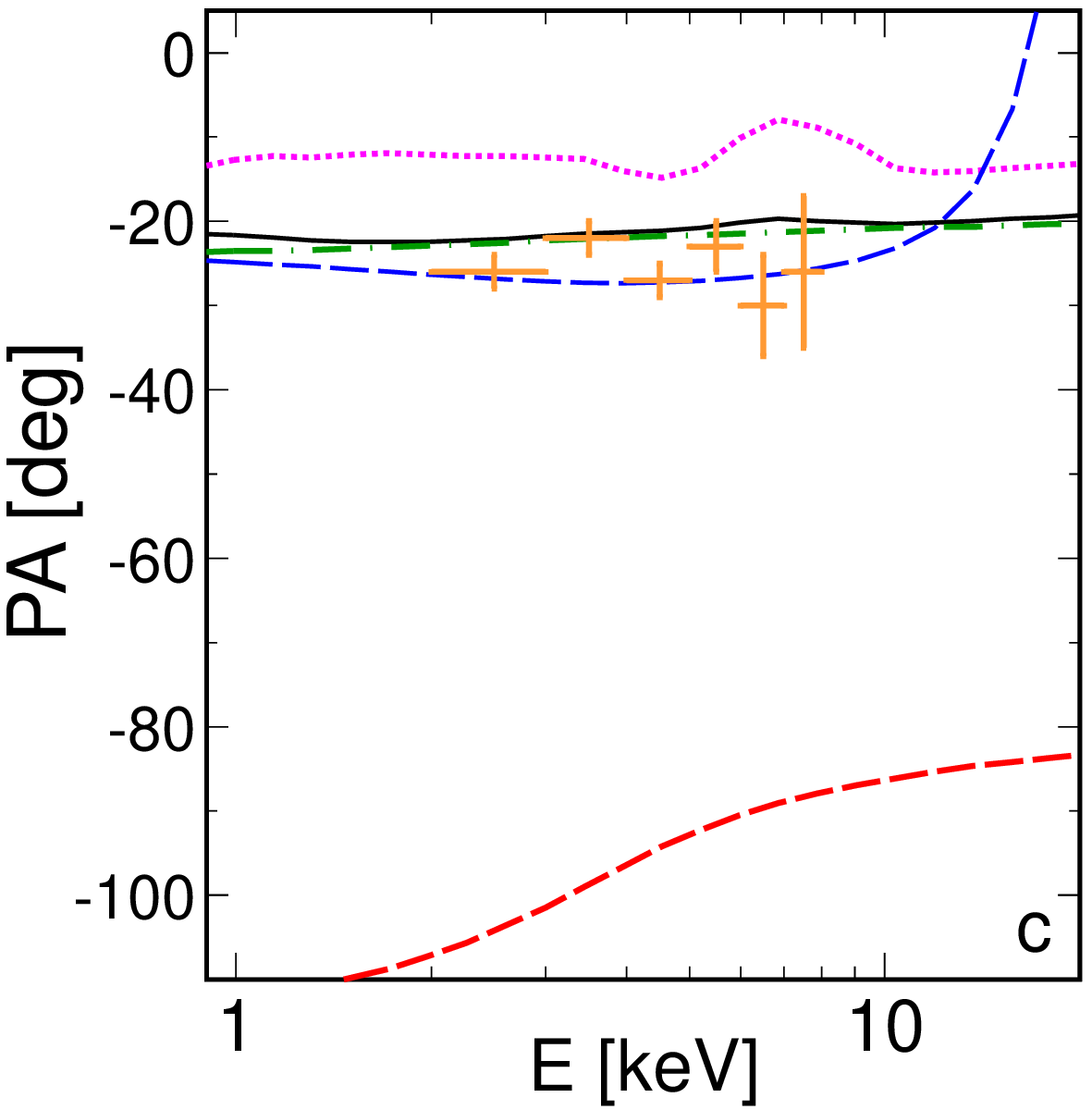}}
\caption{Same as Fig.~\ref{fig:components2} but for model 3 with $a=0$ and including an outflow with $v=0.3c$. The dashed red curves represent the thermal disk spectrum modified by warm-corona Comptonization. 
}
\label{fig:components3}
\end{figure*}

\begin{figure}
\resizebox{\hsize}{!}{\includegraphics{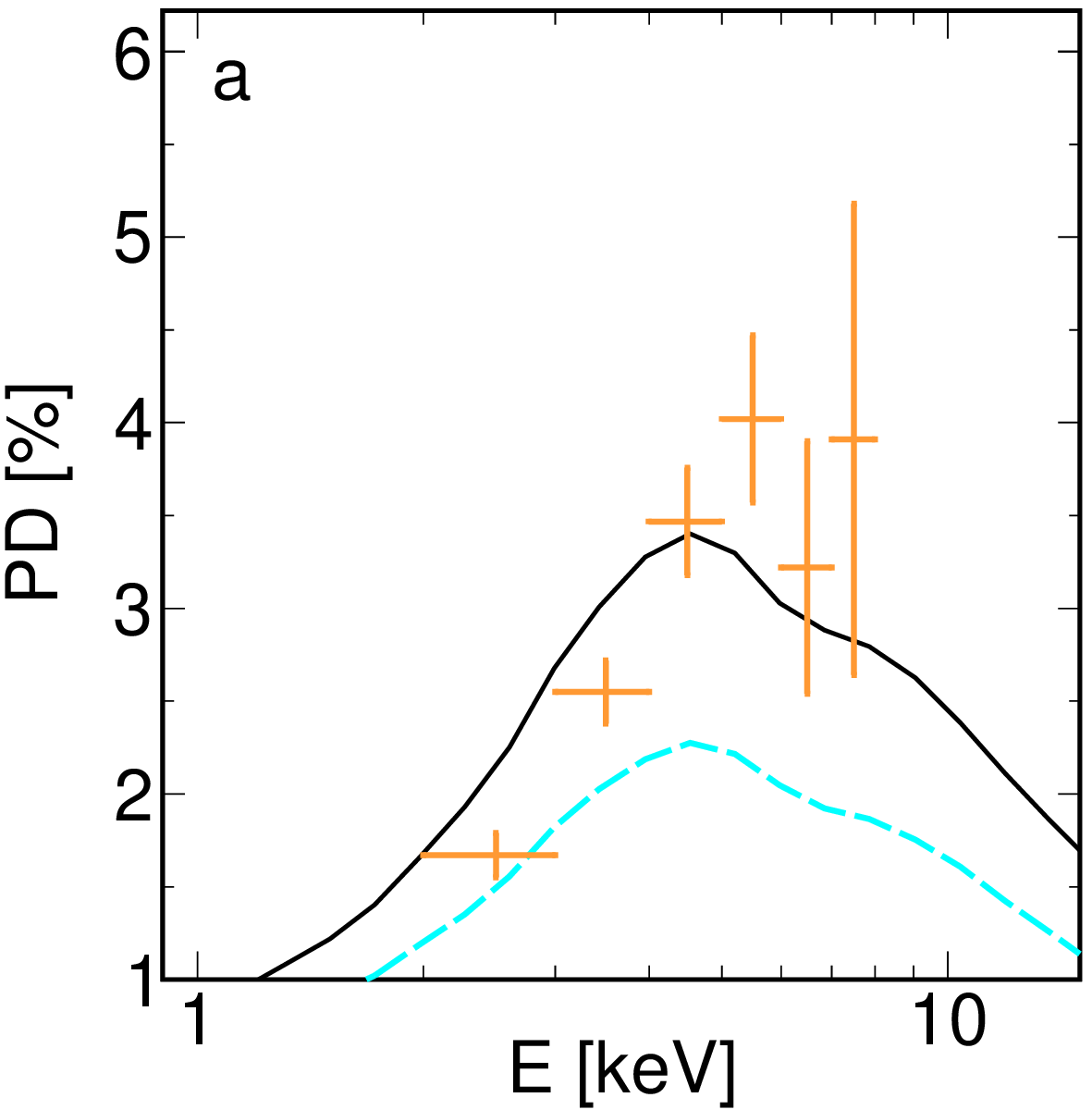}\hspace{0.4cm}\includegraphics{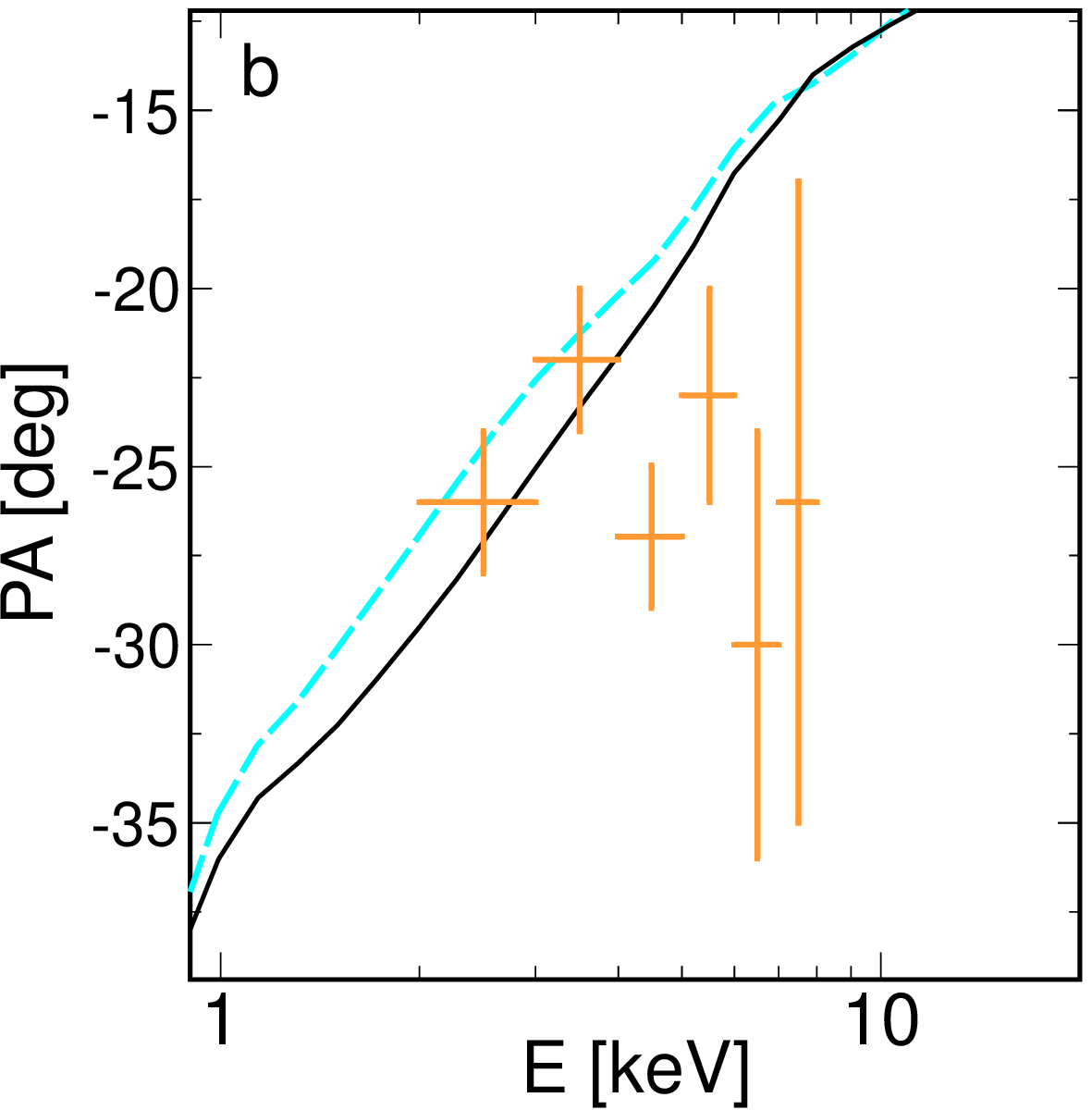}}
\caption{PD (a) and PA (b) of model 2 including an outflow with $v=0.3 c$ for $i=35\degr$ (the fitted value in this model; dashed cyan curves) and $i=39\degr$ (better matching the observed PD; solid black curves). }
\label{fig:iv}
\end{figure}

\subsection{Spectral decomposition}

We first attempted to fit the entire 1--600~keV energy range, but we were unable to fit it above approximately 100 keV, where our models underestimated the data. A likely reason for that is that the nonthermal part of the electron distribution was assumed to be a power law with a sharp cutoff, while a gradual cutoff is more likely \citep[e.g.,][]{1998A&A...333..452K,2008ApJ...681.1725S}.
We thus opted to fit the 1--100~keV energy range only. In all models presented below, the best-fit parameters remain unchanged regardless of whether the OSSE data are included. The only effect of including these data is a modest reduction in the uncertainties of the hybrid plasma parameters in the {\tt comppsc} component. Therefore, our conclusions do not depend on the inclusion of these archival data. We nevertheless present results including OSSE, as these observations have formed the basis of extensive studies on the nonthermal electron component in the soft state, primarily for the 1996 dataset. In this context, it is useful to show that the emission state during the IXPE observations is broadly consistent with the less explored 1994 OSSE results.

We considered models without and with a warm corona (see Fig.~\ref{fig:geometry}; note that the hot corona geometry is simplified in this figure, as discussed in the caption). The model definitions and fitted parameters are given in Tables  \ref{tab:models_def} and \ref{tab:models}, respectively. Figure \ref{fig:mod2} shows the spectrum and residuals for model 2; for models 1 and 3, these are very similar. Figures~\ref{fig:components2}a and \ref{fig:components3}a show the unabsorbed spectral components of models 2 and 3, respectively.

For the case without a warm corona, we obtain a good fit for $a \approx 0.99$. This high spin implies a relatively strong returning component, accounting for $\sim 5$\% of the total flux at $\sim$2 keV where its contribution is greatest (see Fig.~\ref{fig:components2}). 
For the warm-corona case, we also obtain a good fit, but the fitted spin parameter is $a \approx 0$. Consequently, in this case (model 3), the returning component is very weak, contributing less than 1\% to the total flux in the whole energy range. 

Motivated by the results of \cite{2025A&A...698A..37R}, we investigated how our results depend on adopting a lower black hole mass. For $M_{\rm BH} = 17 M_{\odot}$, we find that the change can be compensated by reducing the hardening factor to $\fcolor \simeq 1.3$, yielding a spectral fit of the same quality as that reported in Table~\ref{tab:models}, with only insignificant changes in the remaining parameters. Interestingly, the inclination $i \simeq 35\degr$ fitted to all models is consistent with the binary inclination reported by \cite{2025A&A...698A..37R}. Adopting $i=27\fdg5$ in model 3, as estimated by \cite{2021Sci...371.1046M} for the orbital inclination, yields $\Delta \chi^2 = 17$ relative to our best solution with $i=35\degr$.

\cite{2024ApJ...969L..30S} inferred a crucial role for returning disk radiation under the assumption of a perfectly reflecting disk. This raises the question of whether their conclusions depend on the unity-albedo assumption. We address this issue by comparing the model in which the returning component was computed assuming a unity albedo \citep[as in][model 1]{2024ApJ...969L..30S} with the model employing the {\tt xillverNS} tables (model 2). Both model variants provide an equally good description of the data, without requiring changes in the parameters of the other spectral components. This insensitivity is partly due to the high ionization state of the disk surface (which is required to reproduce the coronal reflection), but primarily due to the very weak contribution of the returning component. Consequently, our conclusions regarding the role of returning radiation are robust against the adopted treatment of its reflection. The results of \citet{2024ApJ...969L..30S} are discussed in more detail in Sect.~\ref{sect:steiner}.

\subsection{Polarization constraints}

Based on the spectral descriptions obtained above, we next computed the corresponding polarization signals, initially neglecting any outflow in the corona. As shown in Fig.~\ref{fig:components2}b, the polarization in model 2 is dominated by the Comptonized component; however, the observed PD is underpredicted by a factor of $\sim2$. The reflected returning component contributes $\lesssim 25$\% to the observed PD around 3 keV, with its contribution rapidly decreasing at higher energies. This weak contribution is readily understood given the small fraction of the total flux provided by this component ($<5$\%).

The coronal radiation undergoes significant, energy-dependent rotation of the PA due to special- and general-relativistic effects, which is reflected in the energy dependence of the net PA predicted in this model (see Fig.~\ref{fig:components2}c). For large values of $a$, the coronal emission contains a strong contribution from small radii ($\la 5R_{\rm g}$), where these PA rotation effects are most pronounced. The polarized component of the Comptonized emission depends on radius through the radial dependence of the seed-photon temperature (see Fig.~\ref{fig:outflow}), and the largest PA rotations are observed at the highest energies (dominated by these small radii). In particular, at the inner edge of the disk, the color temperature reaches $kT_{\rm col} \simeq 0.6$~keV. Consequently, first-order scattering dominates up to the \textit{IXPE} energy range, and the polarized spectrum rises steeply as the emission transitions to domination by higher-order scatterings. 

The effects observed in the PA of the remaining spectral components are comparatively minor, owing to their low contribution to the net polarization. 
Nevertheless, we discuss them here for completeness. 
The returning component is produced primarily by the reflection of photons emitted near the inner edge of the disk ($\sim 1.5 R_{\rm g}$) and subsequently returning to the disk over a wide range of radii. The emission-to-return redshift increases with the radius at which the photons return. 
The returning radiation observed below 10~keV is dominated by reflection from radii $\ga 5 R_{\rm g}$, where the relativistic PA rotation is weak and, consequently, the observed polarization of this component is perpendicular to the disk plane.
In addition, the associated grazing irradiation of the disk surface at these radii leads to reflection toward low-inclination observers at scattering angles $\sim\pi/2$, which results in the relatively strong intrinsic polarization of this component at an $\sim 10$\% level. 
On the other hand, our description of the coronal reflection is phenomenological and involves assumptions that reduce the PD.  
First, we assumed locally isotropic irradiation of the disk surface, which produces a less polarized reflected signal for low-inclination observers than grazing illumination. 
Second, we assumed that the rest-frame reflection spectrum is the same at all radii, which leads to partial depolarization through the mixing of radiation whose PAs undergo different relativistic rotations; for returning radiation, such a depolarizing mixing does not occur because the rest-frame reflection depends on the radius through redshift. 
Also, the mixing of polarization signals with identical rest-frame spectra but differently rotated PA leads to net rotation by $\simeq -5\degr$, which is about the same at all energies (for the fitted $\beta \simeq 2.8$).

Figure~\ref{fig:iv} shows the polarization properties of model 2, which includes an outflow with $v=0.3c$, and illustrates the dependence on inclination, to which the observed PD is highly sensitive. We see that the predicted PA still exhibits a strong energy-dependent rotation, making it inconsistent with the observed PA. Such behavior is in fact expected for any high-spin model in which the emission follows the radial dissipation profile of a \cite{1973blho.conf..343N} disk, specifically, it includes a strong contribution from the innermost few gravitational radii. The mixing of contributions with differently rotated PAs also reduces the net polarization, and for the fitted inclination of $i=35\degr$ the observed PD is still underpredicted. The PD can be matched more closely by adopting a larger inclination, $i=39\degr$; fixing the inclination in model 2 at this value worsens the fit, yielding $\Delta\chi^{2}=9$.

\begin{figure*}
\resizebox{\hsize}{!}{\includegraphics[height=5.2cm]{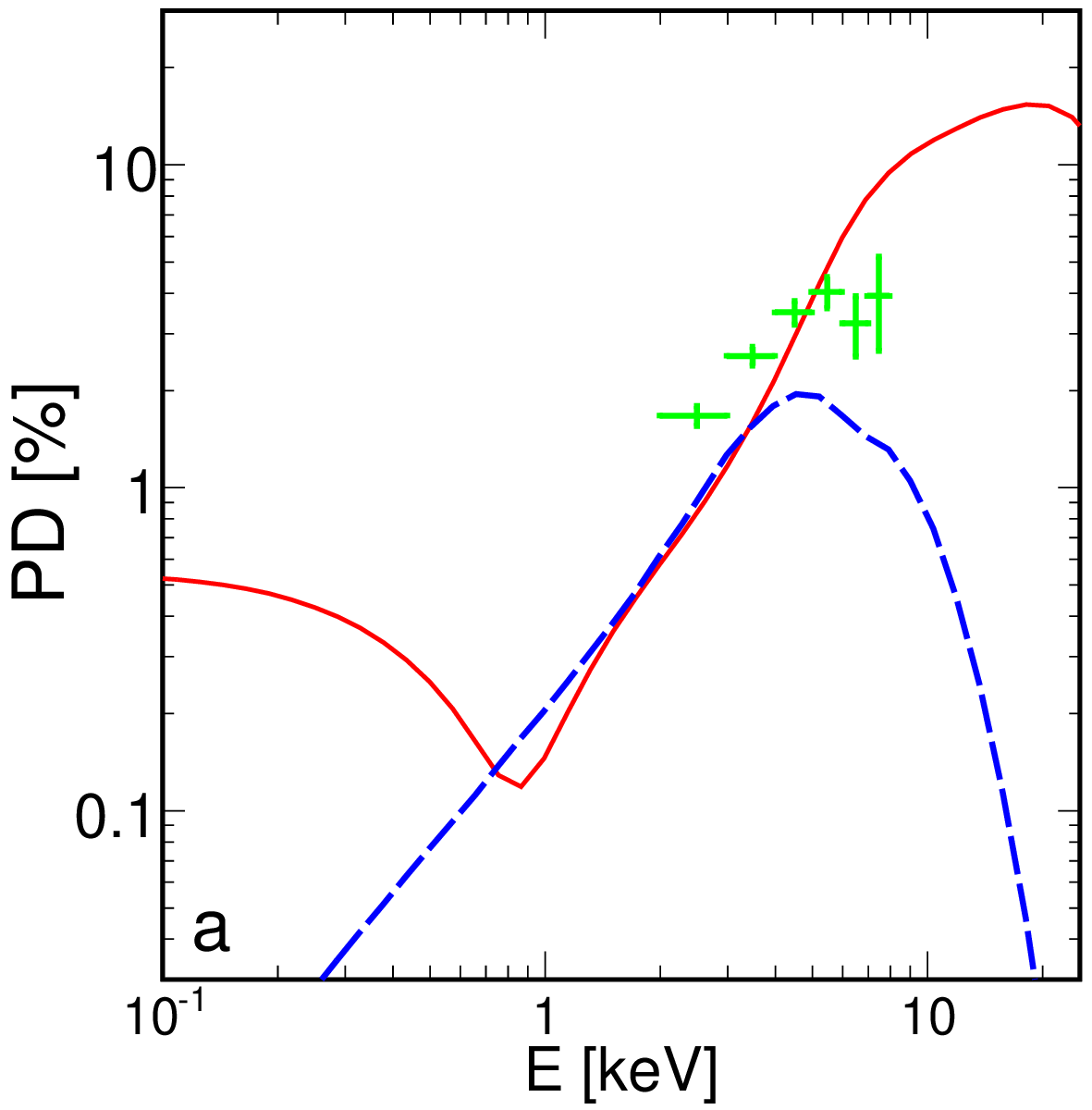} \hspace{0.1cm}
\includegraphics[height=5.2cm]{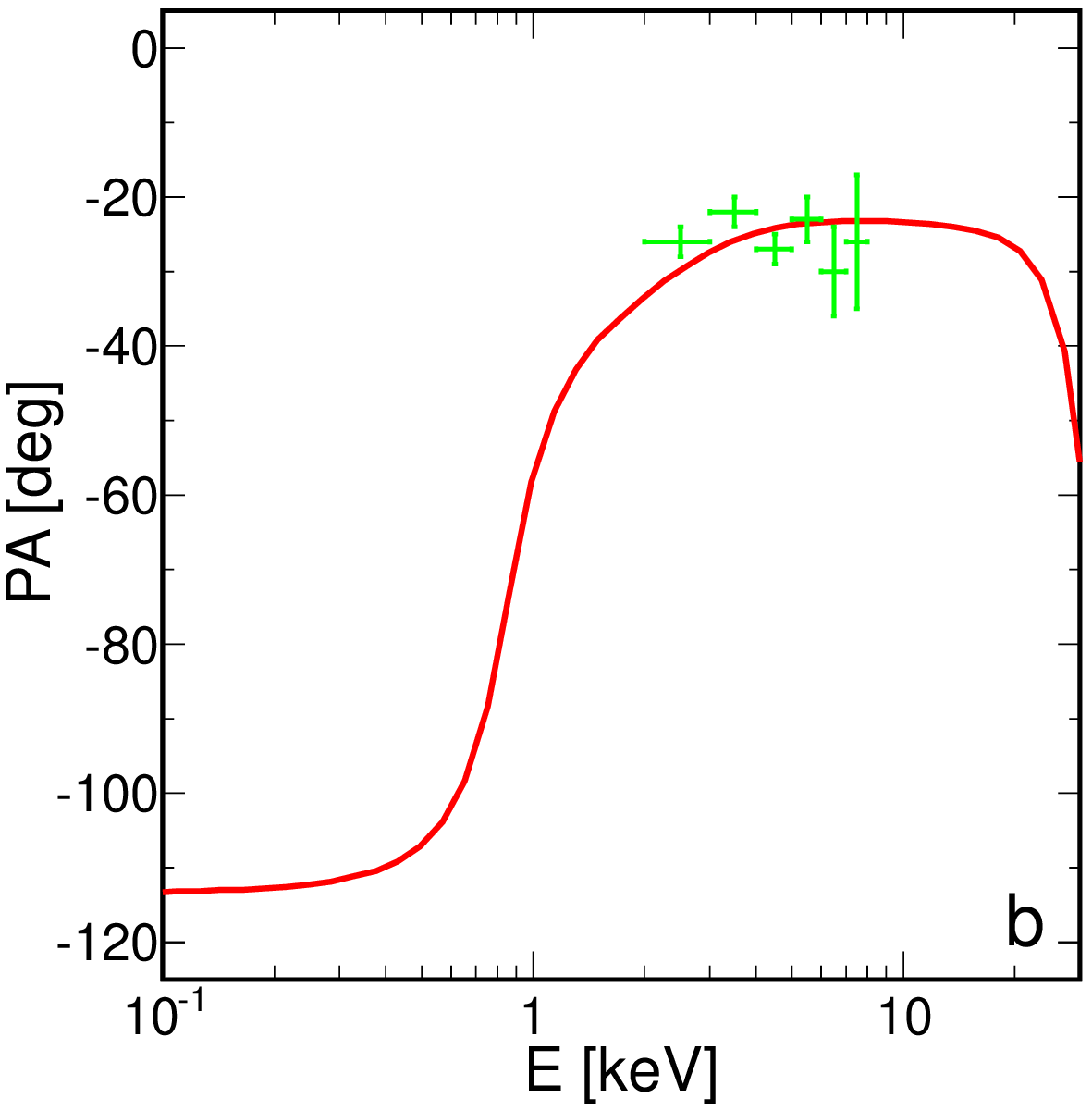}\hspace{0.3cm} \includegraphics[height=5.35cm]{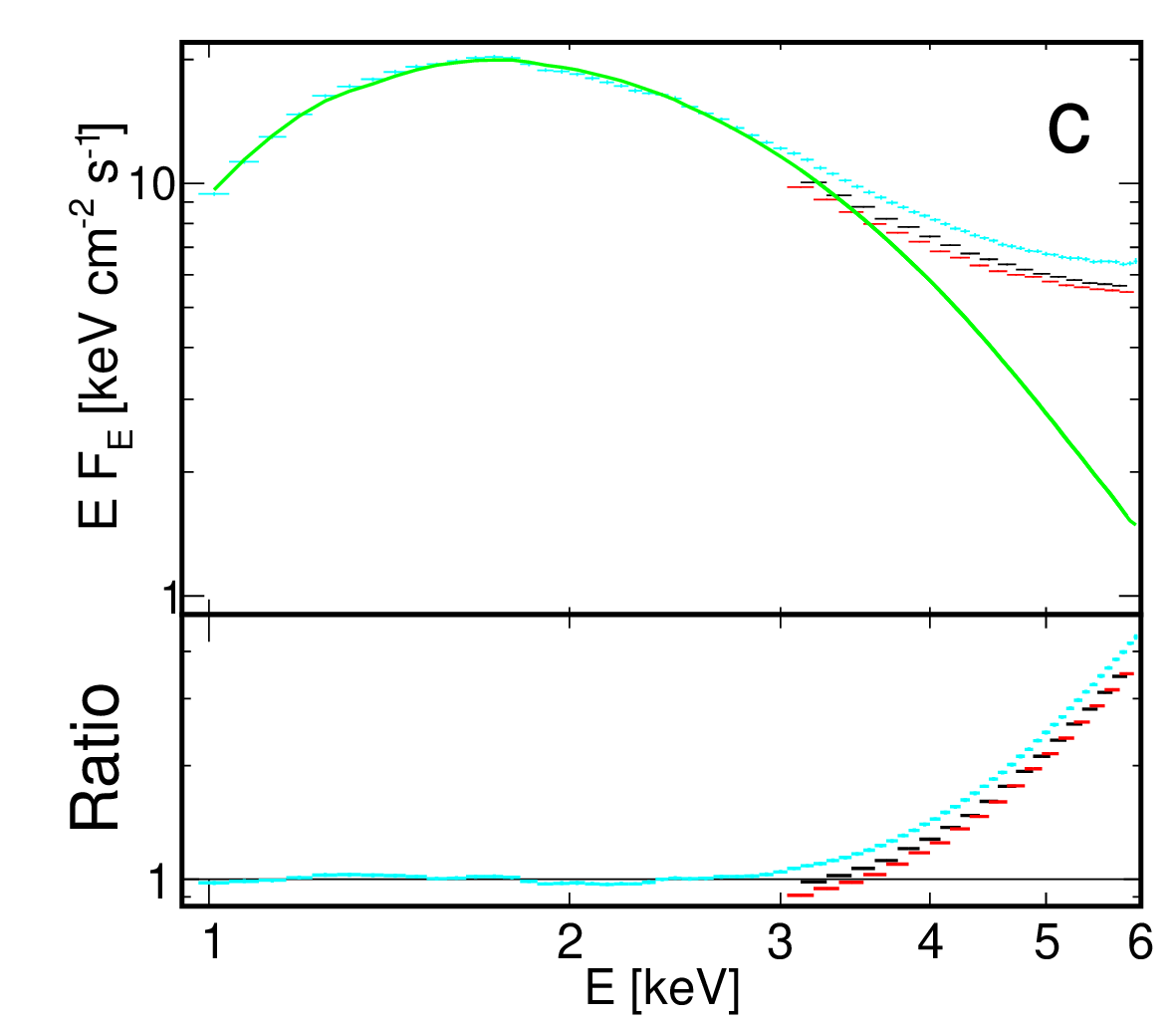}}
\caption{Comparison of model 4 with the spectropolarimetric data on Cyg X-1. 
Panels (a) and (b): Comparison of the polarization properties of model 4 (solid red curves), following the assumptions of the "no-corona" model in \cite{2024ApJ...969L..30S} (i.e., including only the thermal disk), with the soft-state \textit{IXPE} data on Cyg X-1. 
Panel (c): Attempted fit of this model to the {NICER} and \textit{NuSTAR} spectral data. The observed data (cyan, red, and black; as in Fig.~\ref{fig:mod2}) are shown in the top panel, together with the best-fit model (green), while the data-to-model ratio is shown in the bottom panel. The dashed blue curve in panel (a) shows the contribution to the PD of the returning disk component for the parameters used in \cite{2024ApJ...969L..30S}; see Sect.~\ref{sect:steiner}. }
\label{fig:mod4}
\end{figure*}

Figure~\ref{fig:components3} shows the spectral components and polarization for model 3, which includes a warm corona and a black hole with $a \simeq 0$. However, the polarization calculations implemented in \texttt{compps} allow only Planckian seed photons, whereas in model 3 the seed photons follow the spectrum of optically thick Comptonization. We therefore approximated the actual \texttt{thcomp}$\cdot$\texttt{retBB} input with a \texttt{retBB} spectrum using a color-correction factor $\fcolor=2.8$, chosen to reproduce a similar spectral shape. Consequently, the polarization predictions in this case should be regarded as approximate. Apart from the negligible contribution of returning radiation, the relative roles of the individual components are very similar to those in model 2. As in this model, the observed polarization is dominated by Comptonization in the hot corona and, for $v=0$ (and $i=35\degr$), the predicted polarization is weaker than the \textit{IXPE} measurement. However, the coronal emission is now dominated by radii $\gtrsim 10 R_{\rm g}$, where relativistic effects are weak and polarization remains approximately perpendicular to the disk plane. Models including a low-spin black hole therefore generally reproduce the observed PA. The weak PA rotation also implies only weak depolarization due to the mixing of contributions with different PAs. Figure~\ref{fig:components3} shows predictions for a model with $v=0.3c$, which in this case reproduces both the observed PD and PA for $i=35\degr$. In Fig.~\ref{fig:components3} we assumed $a=0$. However, for low values of $a$ the dependence on the exact spin value is weak, and adopting $a=0.2$ (within the uncertainty range of model 3) yields polarization properties very similar to those obtained for $a=0$.

Interestingly, the observed PD shows a hint of a decline in the 6--7 keV energy bin, which may be associated with the contribution of Fe~K$\alpha$ photons at these energies. 
The model predicts a similar decline in PD, although its magnitude is underestimated because our treatment assumes that the entire reflected spectrum is polarized perpendicular to the disk. In contrast, the Fe~K$\alpha$ line should instead be polarized parallel to the disk due to scattering of a fraction of fluorescent photons. 
This effect cannot be consistently captured within reflection models based on the \texttt{xillver} tables.

\subsection{"No-corona" and reflection-dominated models}
\label{sect:steiner}

The importance of returning radiation in producing the net soft-state polarization of Cyg X-1 was previously supported by a model without a Comptonizing corona, discussed in Sect.\ 4 of \citep{2024ApJ...969L..30S} and presented by the dashed orange curves in their Fig.\ 4.
To address this scenario, we considered model 4, constructed under the same assumptions (see Fig.~\ref{fig:geometry}c and the definition in Table~\ref{tab:models}). 
Such a configuration does indeed produce a polarization signal consistent with the \textit{IXPE} measurements (the underlying properties of the returning radiation, resulting in its high polarization and negligible PA rotation at \textit{IXPE} energies, are discussed above; see also Appendix \ref{sect:retbb}), as shown in Fig.~\ref{fig:mod4}. 
However, removing the hot corona, which primarily increases the PD by reducing the total flux, also prevents the model from reproducing the observed high-energy emission. Although \citet{2024ApJ...969L..30S} did not intend this "no-corona" configuration to provide an acceptable fit to the broadband spectrum, it is important to note that its contribution to the observed flux is too weak for the highly polarized returning-radiation reflection component to play a major role in explaining the measured polarization. To quantify this limitation, we compared this configuration directly with the observed spectral data. An attempt to fit this configuration to the NICER and \textit{NuSTAR} data is summarized in Table \ref{tab:models}, and the severe underprediction of the flux above 3 keV is illustrated in Fig.~\ref{fig:mod4}c.

\begin{figure}
\resizebox{\hsize}{!}{\includegraphics[width=0.85\columnwidth]{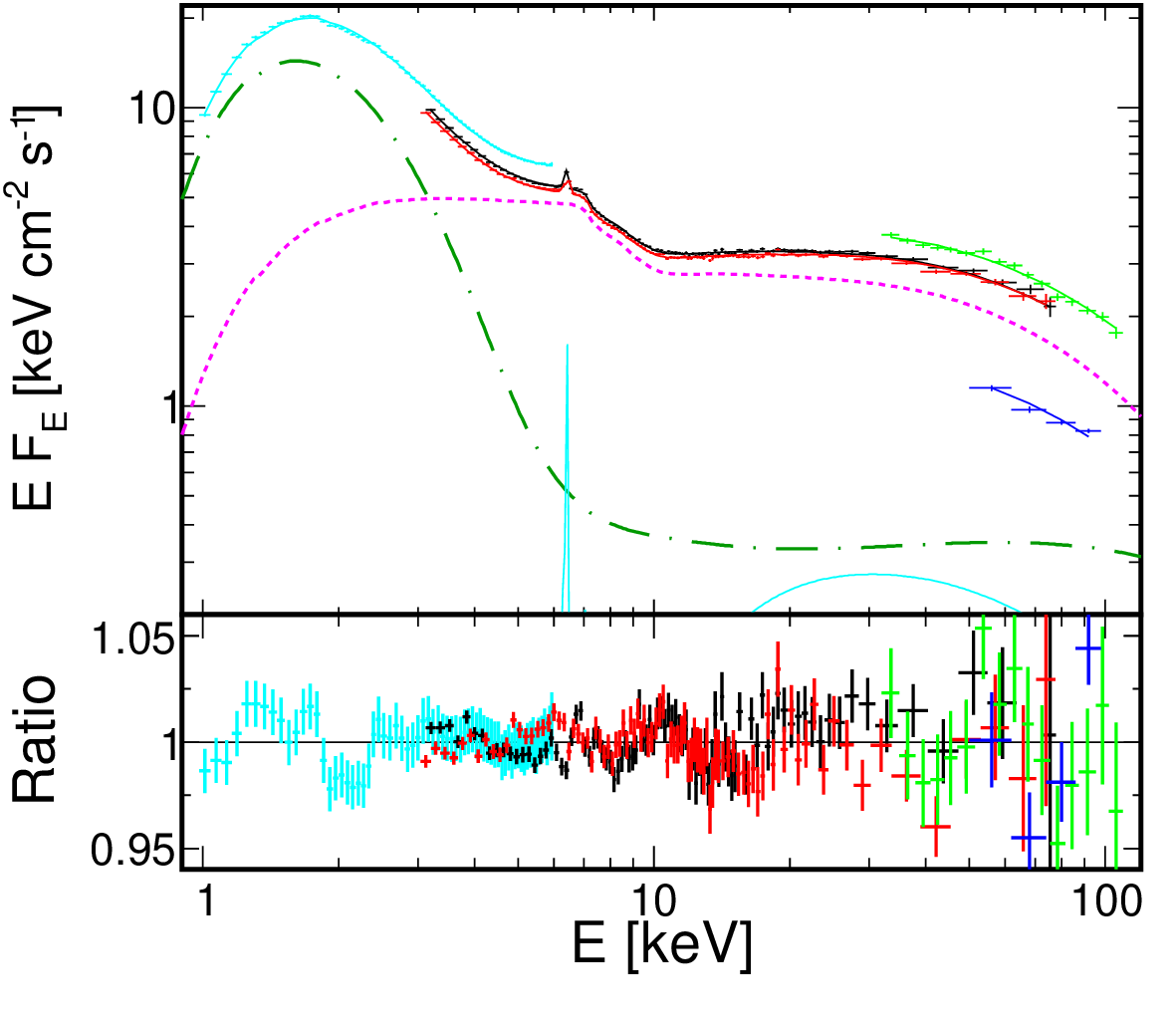}}
\caption{
Same as Fig.~\ref{fig:mod2} but for model 5, i.e., with the $\mathcal{R} \ge 7$ constraint.
}
\label{fig:model5}
\end{figure}

The model used earlier to reproduce both the spectrum and polarization (albeit not constrained by a fit to the observed energy spectrum), included a Comptonized emission and its reflection, the latter strongly exceeding the incident emission, by a factor of $\simeq 7$ at 5 keV \citep[see the blue curves in the right-hand panel in figure 4 in][]{2024ApJ...969L..30S}.
A key assumption was that the disk behaved as a perfect reflector.
To assess whether such a reflection-dominated scenario is compatible with the data 
for the case of a more realistic description of reflection,
we repeated our spectral analysis while retaining the assumptions of model 2 but imposing the constraint $\mathcal{R} \ge 7$. This defines our model 5. The resulting fit provides a substantially worse description of the data, with $\Delta \chi^2 = 101$ (761 compared to 660 in model 2), and clear residuals appear around the Fe K$\alpha$ region, as shown in Fig.~\ref{fig:model5}.

We also considered a phenomenological spectral model that strictly follows the definition of the fully relativistic model in \citet{2024ApJ...969L..30S}, as given in their Sect. 3.2.2. 
In this case, we find parameters similar to those reported in their Table 2, in particular $\mathcal{R} \simeq 0.79$, which again does not support a strong reflection dominance. When applied to our data, this model produces a residual pattern very similar to that shown in Fig.~3 of \citet{2024ApJ...969L..30S} and yields $\chi^2_{\nu} = 1036/644$. 
The spectral models developed here, therefore, provide a statistically better description of the data.

\begin{figure}
\resizebox{\hsize}{!}{\includegraphics[height=5.2cm]{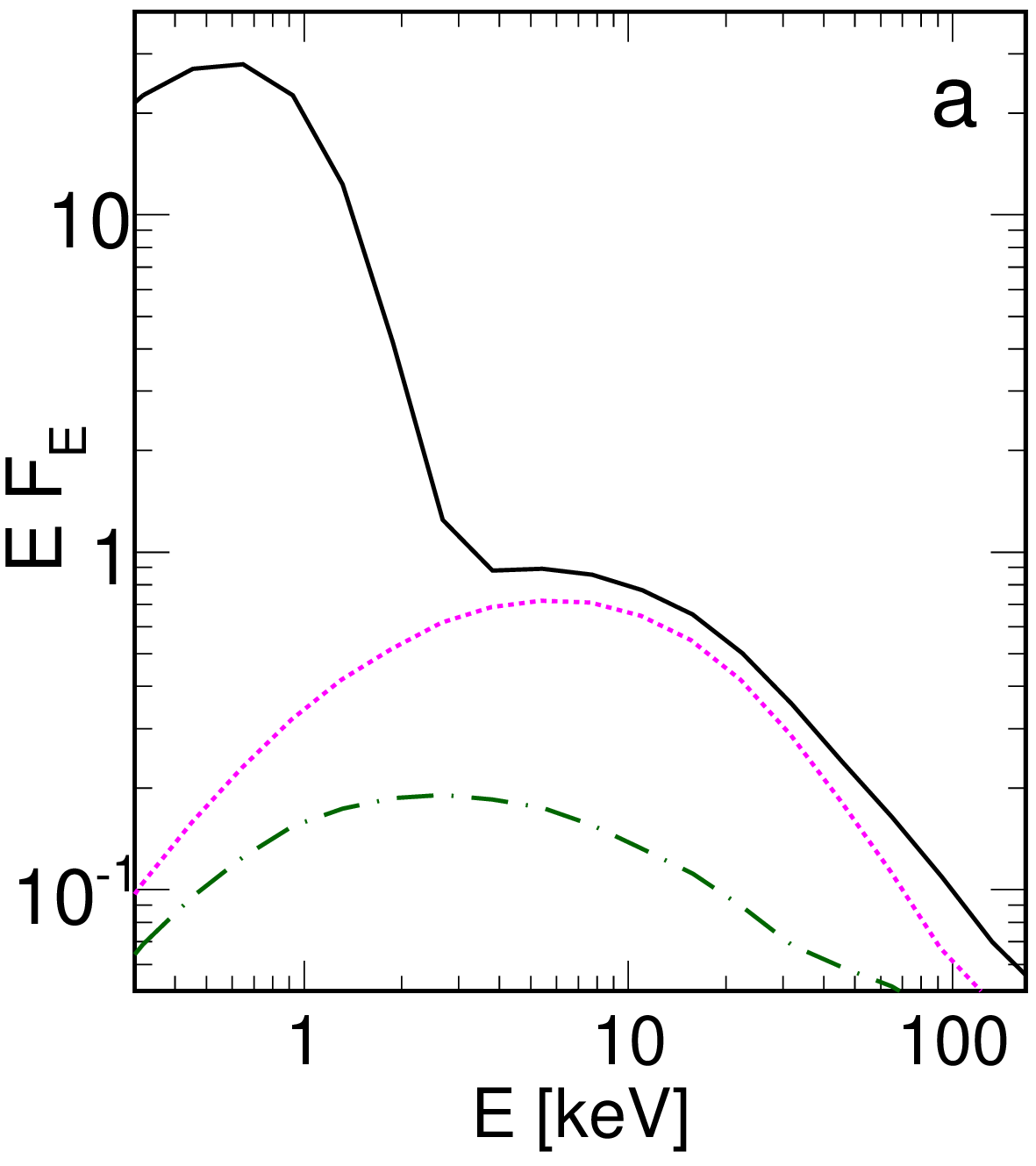}\hspace{0.1cm}\includegraphics[height=5.2cm]{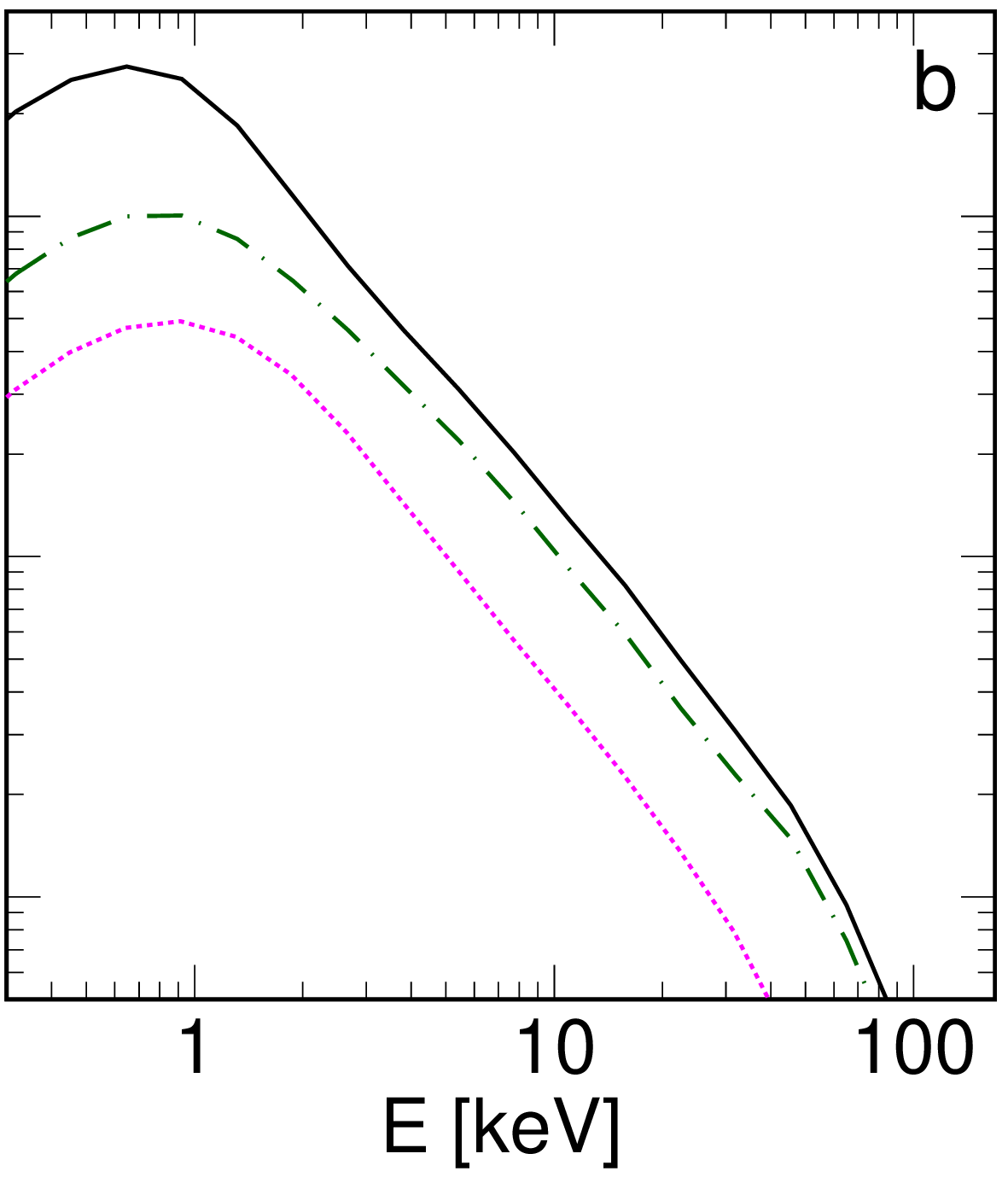}}
\caption{Spectra observed at $i=30\degr$ from a disk-corona system for the accretion-disk parameters corresponding to those in our model 3 and a slab corona, with (a) $\tau=0.007$ and $kT_{\rm e} = 250$ keV and (b) $\tau=0.75$ and $kT_{\rm e} = 25$ keV, computed using the Monte Carlo code of \cite{2005MNRAS.356..913N}. The solid curves show the total spectra, including the thermal disk emission. The dotted magenta curves represent the reflected component, and the dot-dashed green curves show the direct coronal emission. The reflection is calculated assuming a perfectly reflecting disk. For the nonrotating black hole adopted here, light-bending effects are negligible, and the difference between the reflected and direct components arises from the local anisotropies of radiative processes.
}
\label{fig:local}
\end{figure}

Parenthetically, we note that reflection dominance in the physical model \citep[i.e., computed with the \texttt{kerrC} model of][]{2012ApJ...754..133K} presented in \citet{2024ApJ...969L..30S} arises primarily from local anisotropy of Comptonization rather than from light-bending effects. 
The fraction of photons returning to the disk depends on the radial emissivity profile and, even for extremely steep emissivities, remains below 50\% \citep[see, e.g., Fig.~4 in][]{2021ApJ...910...49R}. 
In the \texttt{kerrC} setup, the radial emissivity of the corona closely follows that of the disk. 
Therefore, as in the case of disk emission, general-relativistic light bending and photon return can modify the flux by at most $\sim 10$\% and thus cannot account for reflection dominance by a factor of $\sim 7$.
Comptonization in a slab corona irradiated from one side is intrinsically anisotropic, with the enhancement of the backscattered component increasing as $\tau$ decreases and as the seed-photon energy increases \citep[e.g.,][]{Ghisellini91,1993ApJ...413..507H}. 
For $\tau = 0.007$ \citep[as assumed in][]{2024ApJ...969L..30S} and seed photon temperatures characteristic of a stellar-mass black hole disk, the backscattered flux exceeds the forward-scattered flux by a factor of several (see Fig.~\ref{fig:local}a). This explains the strong reflection dominance in the \cite{2024ApJ...969L..30S} model. By contrast, for $\tau > 0.1$ the local anisotropy becomes weak (see Fig.~\ref{fig:local}b) and arises mainly from the scattering of reflected photons in the corona (an effect included in our spectral modeling).

Finally, we note that the model shown in Fig.~4 of \cite{2024ApJ...969L..30S} was computed using parameters $\dot M = 3.1\times10^{17}$ g s$^{-1}$ and $\fcolor = 1.8$ (Kun Hu, priv.~comm.) for which the disk emission overpredicts the observed spectrum and the peak of the returning component is shifted to $\simeq 3$ keV, compared to $\simeq 2$ keV in our best-fitting model.
When we include a returning component calculated with those same parameters in our model~1, we find that its contribution to the total PD in the 3--4~keV range increases from $\la 25$\% (in our fit-based model) to $\ga 50$\% (see the dotted blue curve in Fig.~\ref{fig:mod4}a). This likely accounts for the difference in our conclusions about the role of this effect.

\subsection{Disk with nonzero thickness}

\begin{figure}
\centering 
\includegraphics[width=0.65\columnwidth]{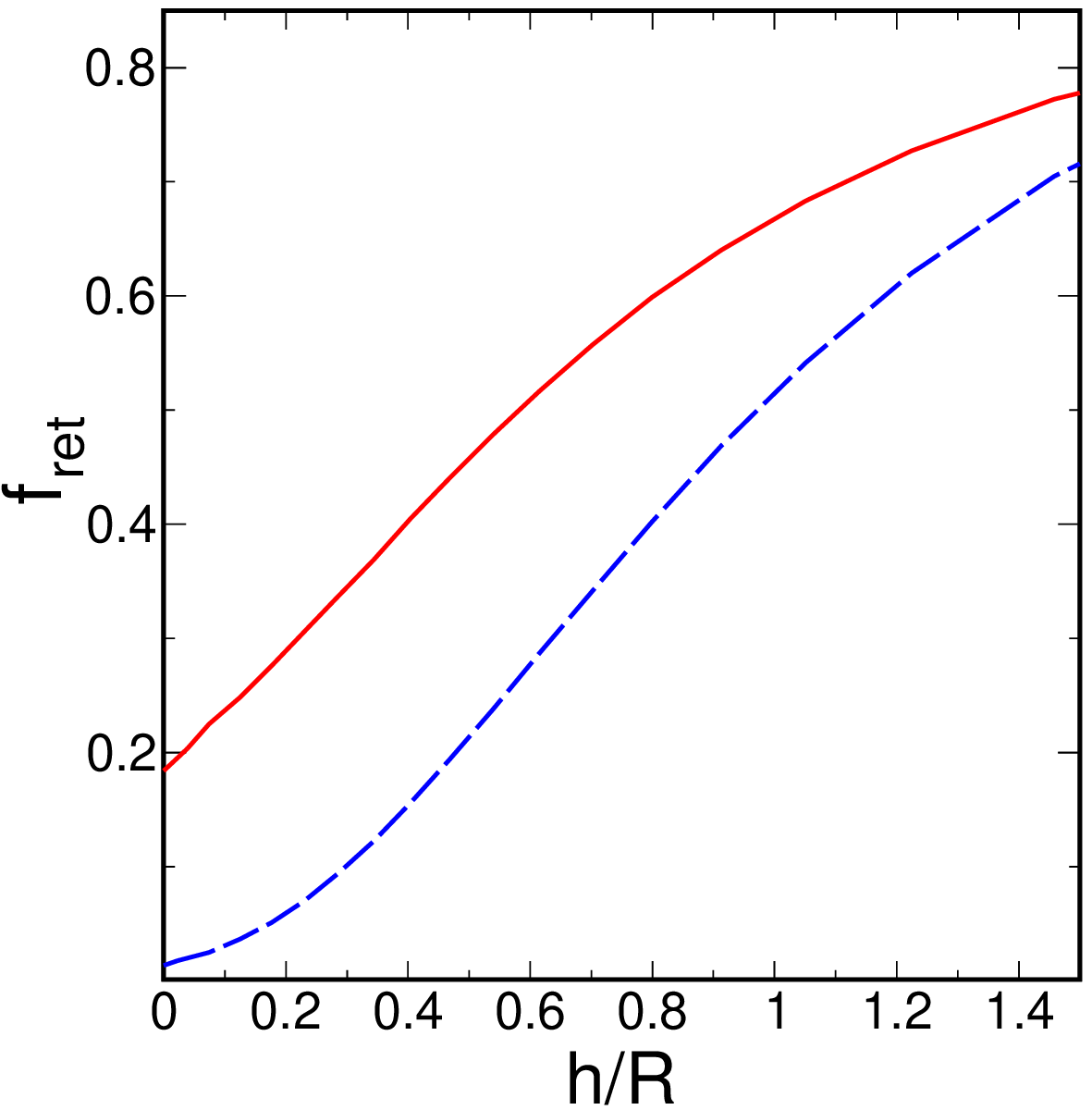}
\caption{Fraction of disk radiation returning to the disk vs. disk aspect ratio for $a=0$ (dashed blue) and $a=0.998$ (solid red). The $h=0$ case corresponds to an infinitesimally thin disk assumed in our main results.
}
\label{fig:flared}
\end{figure}

Our calculations assume an infinitesimally thin accretion disk. If the outer disk is instead flared, the fraction of disk photons intercepted by the disk surface increases. 
In Fig.~\ref{fig:flared} we quantify this effect by showing the fraction of disk radiation returning to the disk as a function of the disk aspect ratio, $h/R$, where $h$ is the half-thickness and $R$ is the cylindrical radius. 
The estimate assumes that the angular distribution of the locally emitted radiation is not significantly altered by flaring, which is justified if the inner disk regions that dominate the luminosity remain only weakly flared. 
The effect shown in Fig.~\ref{fig:flared} arises from purely geometric shielding in flat space-time and is unrelated to general-relativistic light bending. 
It depends only weakly on $a$. 
Consequently, the relative enhancement of returning radiation is larger for small $a$, where the return fraction related to general relativity is comparatively weak. 
We find that for a half-opening angle of $\gtrsim 50\degr$, the reflected returning component could account for the observed polarization around $\simeq 3$~keV. 
However, even in this rather extreme configuration, it cannot reproduce the observed polarization spectrum, as its contribution declines rapidly with increasing photon energy and would produce decreasing polarization in contrast to the observed spectrum. Moreover, such a geometrically thick disk would subtend a solid angle corresponding to $\mathcal{R} \gtrsim 1.5$ at a central X-ray corona, implying a strong quasi-static reflection component. This substantially exceeds the limits on reflection obtained from our spectral analysis. We therefore conclude that this scenario is disfavored.

\section{Summary and discussion}

We performed a comprehensive analysis of the X-ray spectropolarimetric data of Cyg X-1 in the soft state, using convolution models for Comptonization and reflection that self-consistently account for the input photon energy distribution. We also applied our new model, \texttt{retBB}, which describes the disk thermal emission in a manner consistent with \texttt{kerrBB} and additionally incorporates the reflection of returning disk photons. Our study combines two methods for black-hole spin determination: disk-continuum fitting and relativistic reflection modeling. As in a similar analysis of an earlier 2019 Cyg X-1 soft-state observation  \citep{2024ApJ...967L...9Z}, we find that the inferred black hole spin strongly depends on the adopted description of the thermal component. When a warm, optically thick layer Comptonizing the disk radiation is included, the spectral fit is consistent with a low spin value. At the fitted electron temperature of $\simeq 0.5$ keV, the warm layer is not fully ionized; therefore, the relativistic reflection inferred in our analysis may originate in this layer. If the warm layer is neglected, the spectral fit instead favors a near-extreme spin value.

For all our spectral decompositions, we calculated the expected polarization and find that, in our Cyg X-1 model, the polarization signal is dominated by the Comptonized component. 
However, the predicted PD is too low if the Comptonizing plasma corotates with the disk but has no bulk motion perpendicular to the disk plane. 
The predicted PD becomes consistent with the \textit{IXPE} measurement if the plasma undergoes an outflow with a velocity of $\simeq 0.3c$. 
Such an outflow also naturally explains the relatively weak relativistic reflection inferred from our spectral modeling, since the Doppler beaming away from the disk for $v = 0.3c$ corresponds to the reflection fraction ${\cal R} \simeq 0.3$ \citep[see, e.g., Fig.~1 of ][]{1999ApJ...510L.123B} fitted in our models. Physical mechanisms that may drive this type of coronal outflow were proposed in \citet{1999ApJ...510L.123B,2017ApJ...850..141B}, who attributed the coronal emission to magnetic flares dominated by $e^\pm$ pairs that are accelerated away from the disk by radiation pressure. In this scenario, electrons acquire their velocity very close to the disk surface. Indeed, assuming that the acceleration is driven by the local disk radiation pressure and neglecting Compton drag, we estimate that, at radii where most of the coronal emission is produced, the height ($h_{\rm acc}$) over which the pair plasma can be accelerated to $v = 0.3c$ is $h_{\rm acc} \sim 0.01R$ for our best-fit model with $a \simeq 0.99$, and $h_{\rm acc} \sim 0.1R$ for $a \simeq 0$. This picture, in which coronal dissipation occurs in localized blobs, is also consistent with our inferred covering factors, indicating that less than 50\% of the disk surface is covered by hot plasma.

While the spectral fits do not allow us to favor any particular spin value, the polarization measurements can provide a discriminant. 
Radiation from a disk-corona system around a rapidly rotating black hole exhibits a strong relativistic rotation of the PA, making it inconsistent with the observed polarization along the jet direction and thus presumably perpendicular to the disk plane. 
Such rotation is expected whenever a significant fraction of the radiation originates within the central $\simeq 5 R_{\rm g}$. 
For a slowly rotating black hole, the lack of such a contribution naturally follows from the standard assumption of weak energy dissipation within the ISCO. 
Therefore, the observed polarization favors a low spin value.

We emphasize, however, that the assumption that the seed photons originate from the disk emission is crucial for the predicted polarization, as the photon directionality strongly affects the resulting polarization signal. While this assumption is well justified in the soft state, the polarization of the Comptonized component would differ if the seed photon population were instead dominated by internal emission within the hot corona.

Our conclusions differ from those of \citet{2024ApJ...969L..30S} regarding the role of returning radiation. 
In particular, we find only a minor contribution of returning disk radiation even for high spin, in contrast to \citet{2024ApJ...969L..30S}, who indicated this effect to be crucial. 
While the intrinsic polarization of the reflected returning disk component is indeed consistent with the observed PD and PA, its flux contribution is too small to play a significant role in explaining the measured signal. 
This discrepancy appears to arise from the fact that the polarization model presented in \citet{2024ApJ...969L..30S} was not strictly constrained by a spectral fit.
We also do not confirm a significant role of light bending in producing the strong reflection dominance over the direct coronal emission in their polarization model. 
For the extremely low optical depth assumed in that setup, reflection dominance is primarily a consequence of local anisotropy of Comptonization rather than the effects of general relativity. 
Since the reflection component arises predominantly from local disk irradiation, its PD should be relatively low (compared to the $\sim 10$\% polarization of returning disk-radiation reflection, which arises from grazing irradiation of distant regions of the disk).

On the other hand, our calculations of the polarization of the coronal reflection involve assumptions inherent to phenomenological reflection models, namely a radius-independent local reflection spectrum and locally isotropic disk irradiation. 
Both suppress the predicted PD, and relaxing them would likely increase it moderately. 
Nevertheless, reflection cannot dominate the observed polarization, as spectral constraints require that its flux contribution remain relatively small.

An additional argument against attributing the observed polarization to reflection comes from the similarity of polarization signals across soft-state measurements \citep[PA constant with energy and PD increasing with energy;][]{2025A&A...701A.115K}. 
This uniform behavior suggests a common physical mechanism operating in different sources. 
A reflection-dominated origin would generally imply strong reflection features in the energy spectrum. 
However, some sources observed in the soft state, such as 4U~1630$-$47, exhibit largely featureless spectra \citep{2024ApJ...964...77R}. 
This favors Comptonization as the primary mechanism responsible for the observed polarization.

Finally, although our spectral fits include a nonthermal electron component contributing to the high-energy tail, our polarization calculations account only for the thermal electrons, due to current limitations in the \texttt{compps} model used here. 
In the fitted hybrid Comptonization models, the nonthermal tail contributes approximately 13\% of the Compton-component energy flux in the 2--8 keV range. 
Its effect on the net polarization is therefore expected to be small. 
Consistent with this expectation, \citet{2017ApJ...850...14B} find that the nonthermal electron component has a negligible impact on the polarization properties of Comptonized emission in a hybrid plasma with parameters similar to those inferred from our spectral modeling.

\begin{acknowledgements}
We thank Henric Krawczynski and Kun Hu for discussions and Jack Steiner for providing the reduced NICER data. 
We acknowledge support from the Polish National Science Center grants 2023/50/A/ST9/00527, 2019/35/B/ST9/03944 and 2023/48/Q/ST9/00138. 
A.V. and J.P. are supported by the Research Council of Finland grants 355672, 372881, and the Centre of Excellence in Neutron-Star Physics (grant 374064).
Nordita is supported in part by NordForsk. P.L. was partially supported by a program ‘Regional Excellence Initiative’ (project no. RID/SP/0050/2024/1) of the Polish Ministry of Science. AS acknowledges support from the Jenny and Antti Wihuri Foundation (grant no. 00240331).
\end{acknowledgements}

\bibliographystyle{aa} 
\bibliography{references} 

\begin{appendix}
\nolinenumbers

\section{retBB}
\label{sect:retbb}

\begin{figure*}
\centering
\includegraphics[height=5.cm]{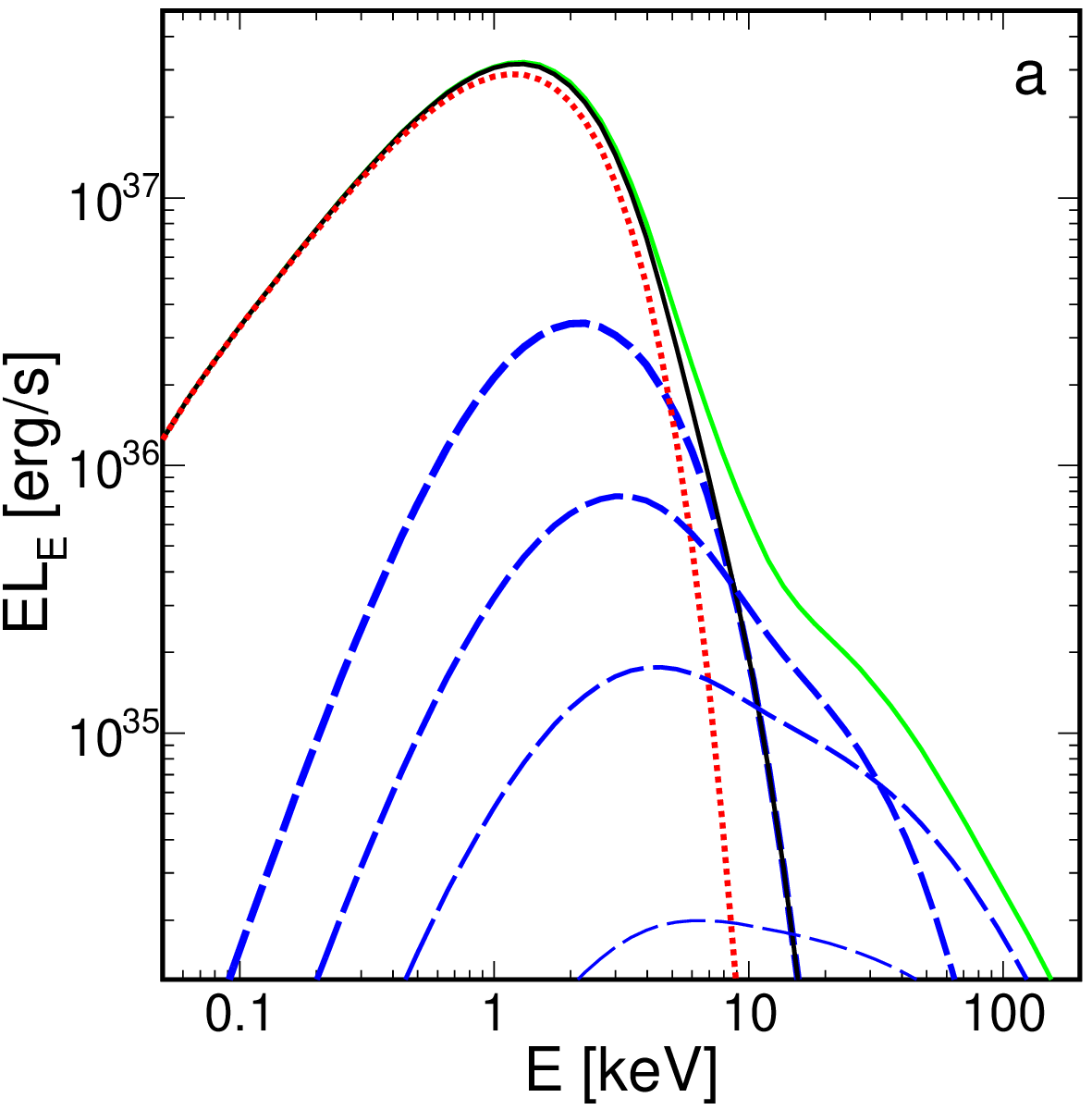} \hspace{0.1cm} \includegraphics[height=5.cm]{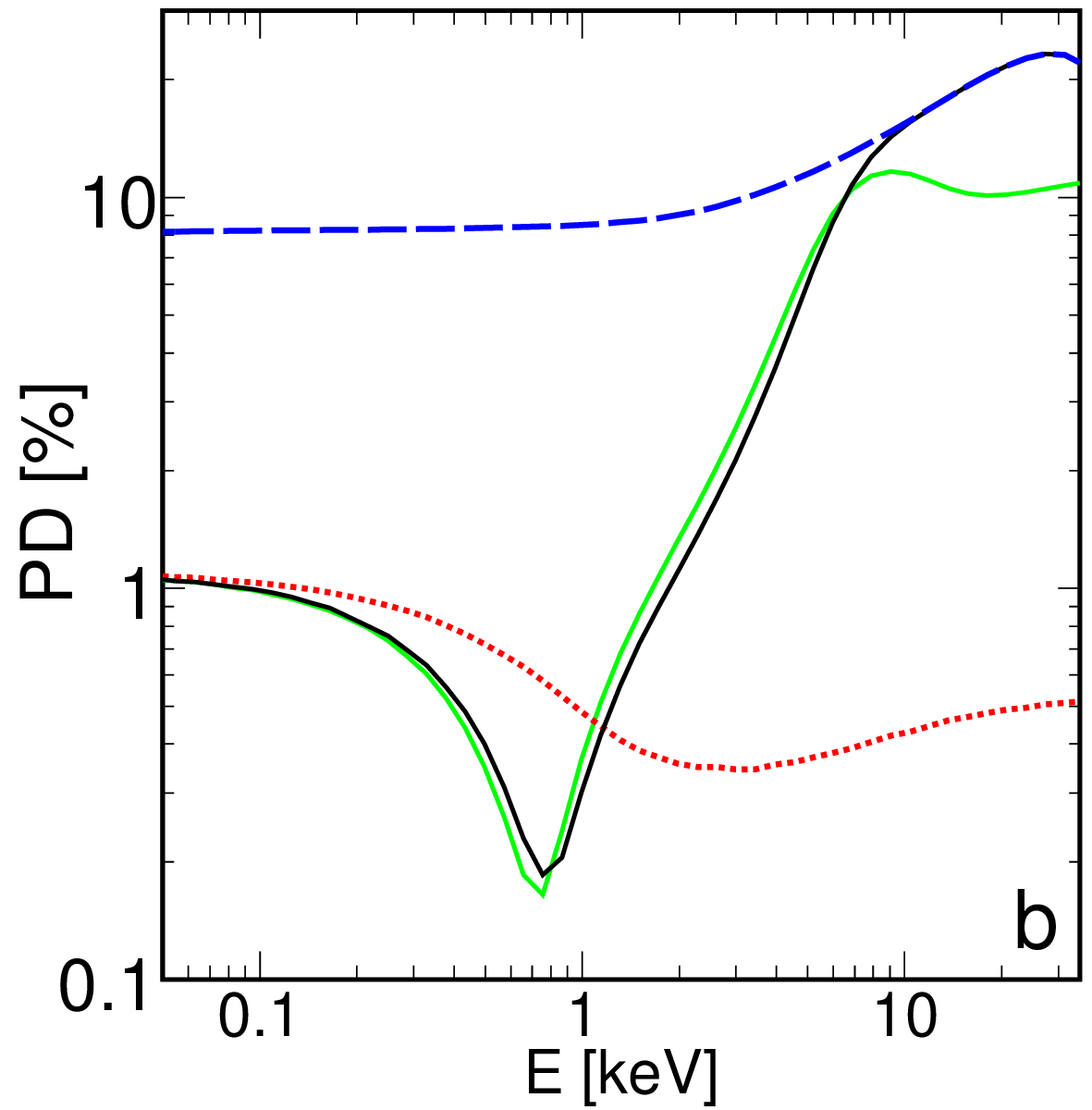} \hspace{0.1cm} \includegraphics[height=5.cm]{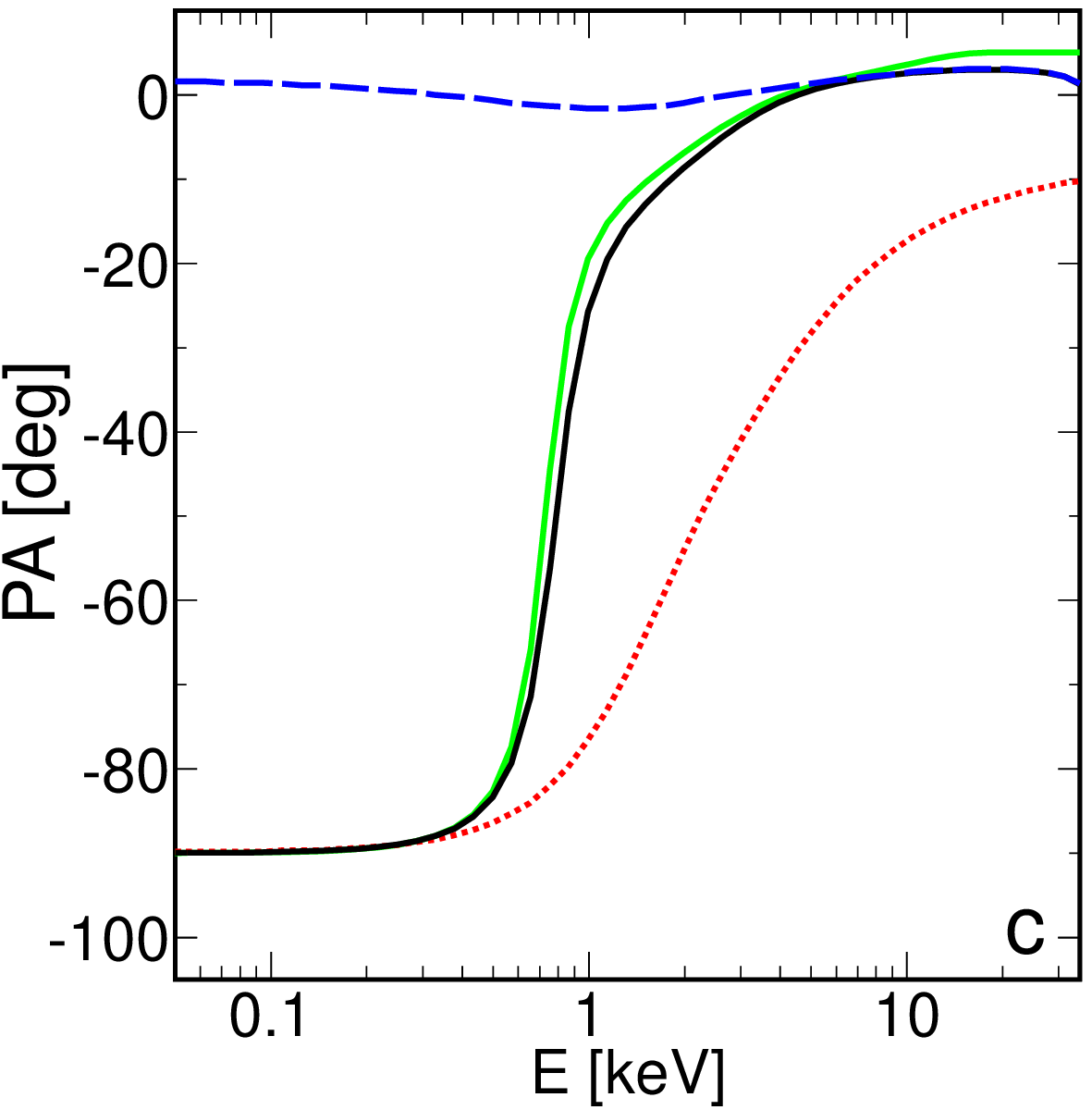}
\caption{Observed spectrum, PD, and PA predicted by \texttt{retBB} for $a = 0.998$, $M=21.2 M_\odot$, $\dot M = 2.5 \times 10^{17}$ g\,s$^{-1}$, $i = 45\degr$ and $\fcolor = 1.5$.
(a) The red dotted curve shows the spectrum of the direct disk emission. The blue dashed curves from top to bottom show spectra of the first to fourth orders of elastic reflection of the returning radiation. The solid black and green curves show the total spectrum, including single- and multiple-scattering (up to fourth order) of the returning radiation, respectively. (b) and (c) PD and PA, including a single (black solid) or multiple (up to the fourth order;  green solid) reflection of returning radiation; blue dashed and red dotted curves show contributions of the first order reflection and direct disk radiation, respectively.
}
\label{fig:multi}
\end{figure*}

\begin{figure*}
\centering
\includegraphics[height=5.cm]{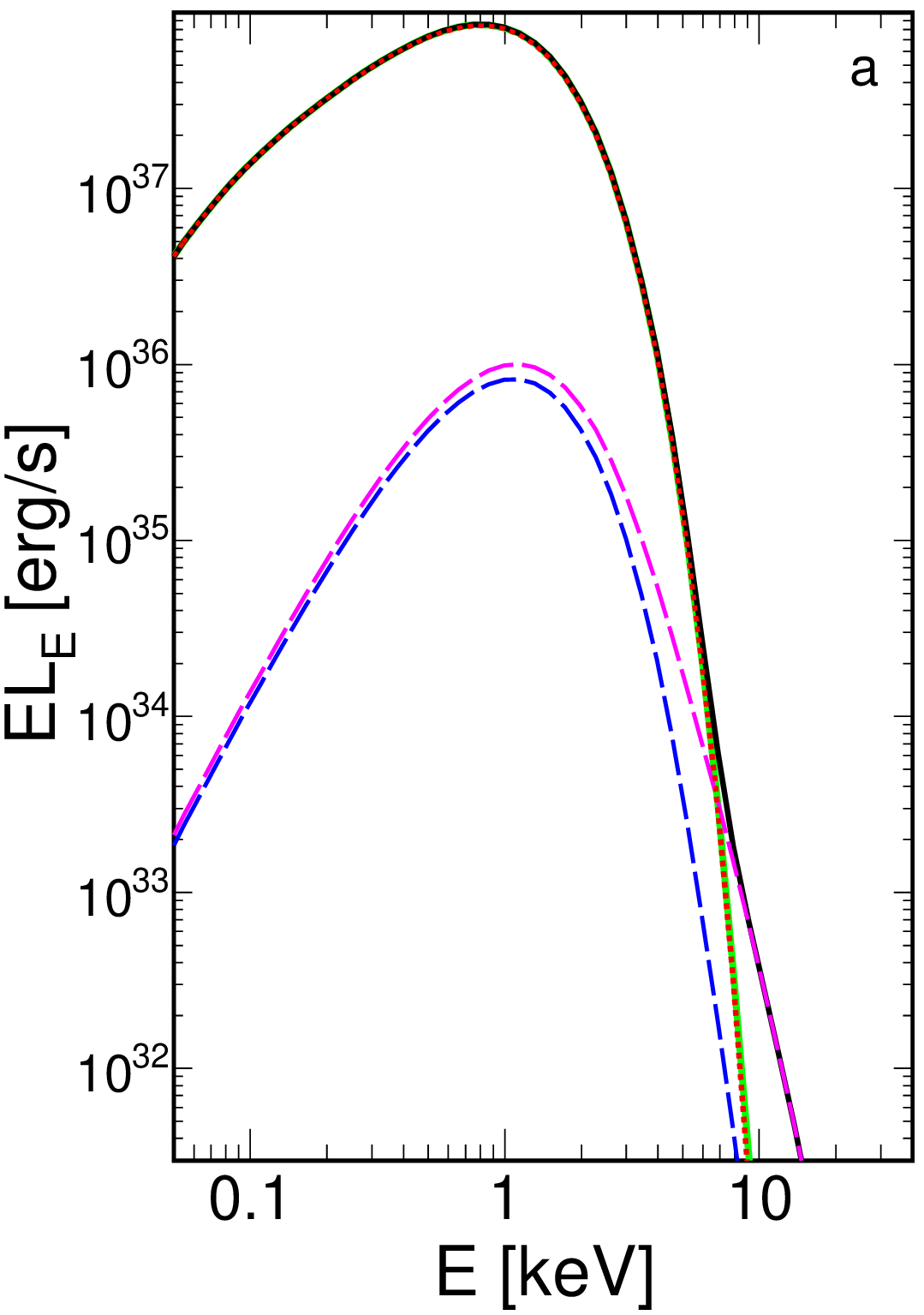} \hspace{0.1cm} \includegraphics[height=5.cm]{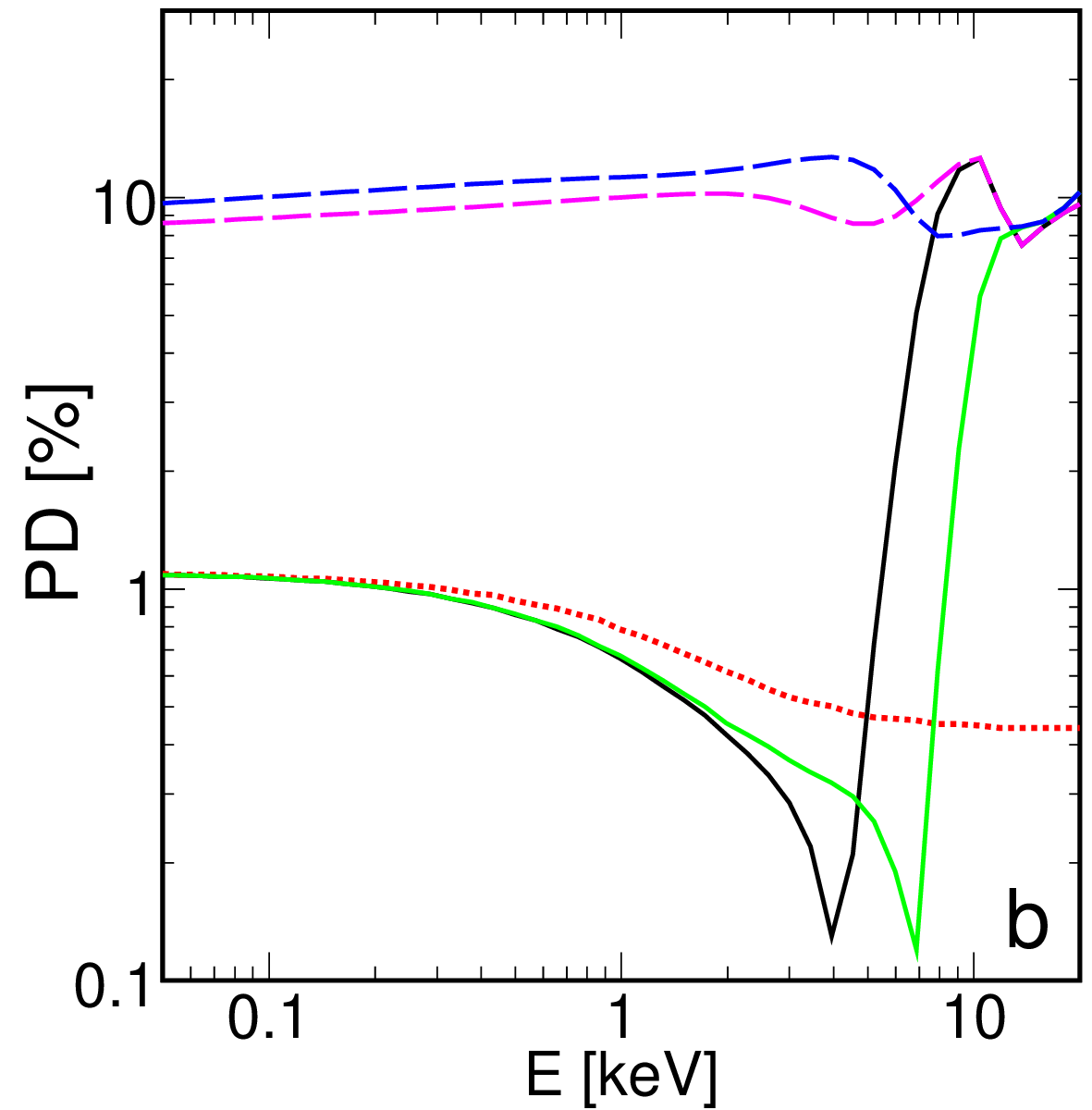}\hspace{0.1cm} \includegraphics[height=5.cm]{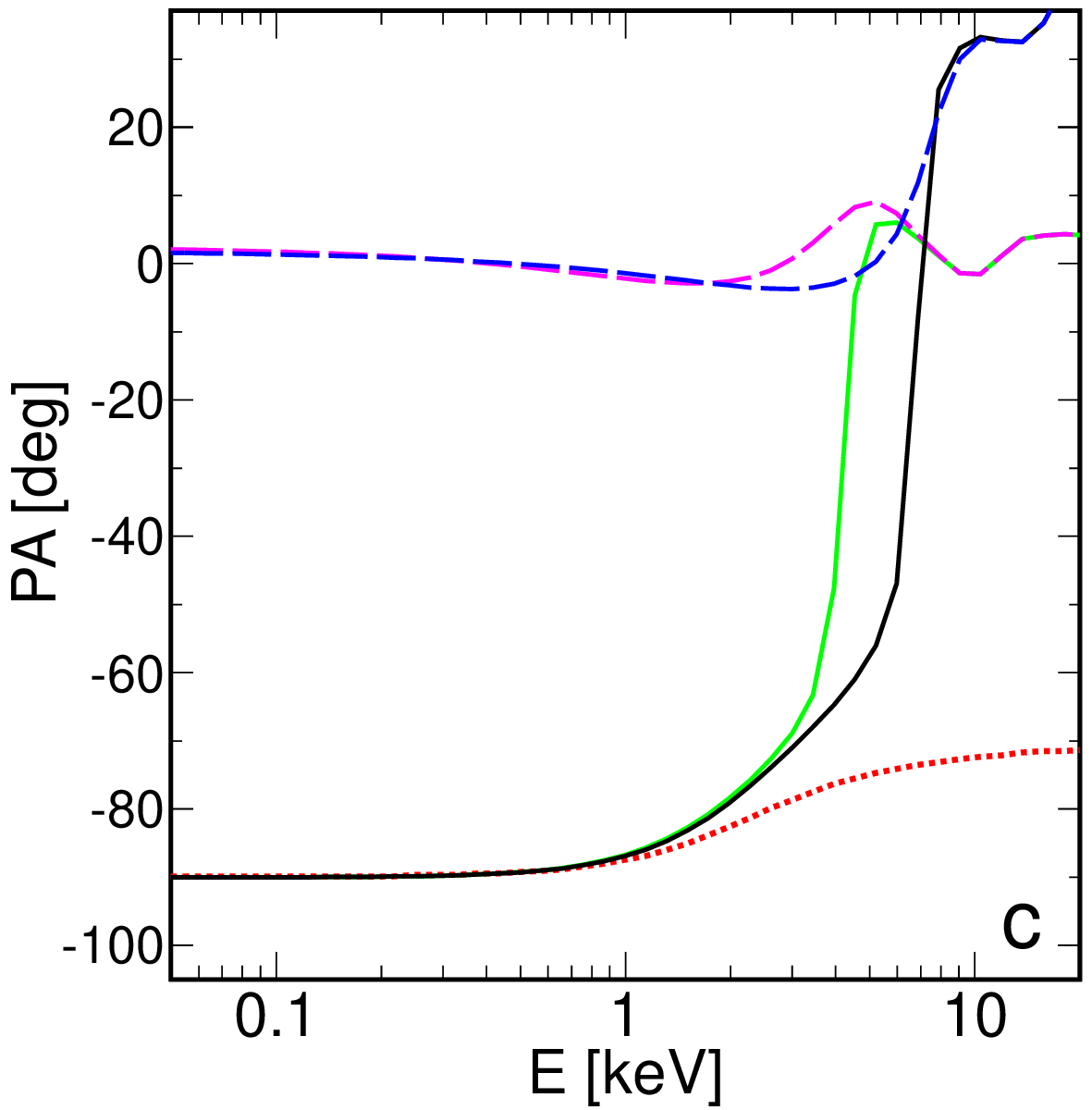}
\caption{Observed spectrum, PD, and PA predicted by \texttt{retBB} for $a = 0$, $M=21.2 M_\odot$, $\dot M = 2.5 \times 10^{18}$ g\,s$^{-1}$, $i = 45\degr$ and $\fcolor = 1.5$.
(a) The red dotted curve shows the spectrum of the direct disk emission. The blue and magenta dashed curves show the elastic reflection of the returning radiation, excluding and including, respectively, the reflection inside ISCO. 
The spectra include multiple reflections; however, contributions beyond the first order are negligible. The solid black and green curves show the total spectrum without and with reflection inside the ISCO, respectively.
(b) and (c) The corresponding PD and PA, the color scheme follows panel (a).
}
\label{fig:isco}
\end{figure*}

Our new model, \texttt{retBB}, extends the \texttt{reflkerr} family of models \citep{2019MNRAS.485.2942N}, by incorporating the reflection of the returning blackbody photons and accounting for the polarization of the reflected and direct emission from an accretion disk. We assume that a flat accretion disk is located in the equatorial plane of a Kerr black hole characterized by the dimensionless angular momentum $a$ and mass $M$. The inclination angle of the line of sight to the symmetry axis is given by $i$. All \texttt{retBB} results presented in this work (except for the green dotted curve in Fig.~\ref{fig:kerrc}) are for the outer radius of $R_{\rm out} = 10^3 R_{\rm g}$. 
We use the formula of  \cite{1974ApJ...191..499P}  for the local radiative flux from a disk accreting at a rate $\dot M$ for a zero-stress inner boundary condition at the ISCO, $\mathcal{F}$,  which determines the local effective temperature $T_{\rm eff} = (\mathcal{F}/{\sigma_{\rm SB}})^{1/4}$, where $\sigma_{\rm SB}$ is the Stefan-Boltzmann constant. 
We approximate the local spectrum of the quasi-thermal emission by a diluted blackbody, with the spectral intensity $I_\nu (\mu)= I_{\rm limb}(\mu) \fcolor^{-4} B_{\nu}(\fcolor T_{\rm eff})$, where $B_\nu$ is the Planck function, $\mu = \cos \theta$, $\theta$ is the emission angle with respect to the normal to the disk measured in the disk rest frame, and $I_{\rm limb}$ is the limb darkening/brightening factor for a plane-parallel, electron scattering-dominated atmosphere tabulated as a function of $\mu$ in Table XXIV in \citet{1960ratr.book.....C}. 
For the local polarization of the thermal disk emission, we use the PD given in the same Table XXIV; it ranges from $\mathrm{PD}=0$  to $\simeq 12$\% for photons emitted normal and parallel, respectively, to the disk surface.

The flux of radiation seen by a distant observer is computed by means of a photon transfer function, ${\cal T}$, 
\begin{equation}
F(E_{\rm obs},i) \! =\!\! 
 \!\int\! {\cal T}(a, r, \theta, \phi, i, g) \, I\left(\frac{E_{\rm obs}}{g}, r, \theta\right) \, 2 \pi r \, {\rm d}r \, {\rm d}\phi \,  {\rm d}\mu \, {\rm d}g, 
\label{eq:ref}
\end{equation}
where $g = E_{\rm obs} / E_{\rm disk}$. 
We assume local azimuthal symmetry of the thermal emission, so the intensity $I$ in Eq.~\ref{eq:ref} does not depend on the azimuthal emission angle in the disk frame, $\phi$. However, we use a $\phi$-dependent transfer function because it is also applied to the reflected radiation, for which azimuthal symmetry does not necessarily hold.
The observed Stokes parameters are computed using analogous transfer functions, ${\cal T}_{\rm X}$ and ${\cal T}_{\rm Y}$.
The transfer functions ${\cal T}$, ${\cal T}_{\rm X}$ and ${\cal T}_{\rm Y}$ are defined in the same way as in \cite{1990MNRAS.242..560L} and treat the effects of special and general relativity on the observed total and polarized radiation. 
We construct them by following a large number of photon trajectories originating in the disk until they are captured by the black hole, return to the disk plane, or reach a distant observer. 
We compute the photon trajectories using the code employed for the transfer functions of the other \texttt{reflkerr} models, to which we have added a description of the photon beam polarization, based on the methodology developed in \cite{1980ApJ...235..224C}, \cite{1997PhDT.........2A} and \cite{2016MNRAS.462..115D}. 
In it, the parallel transport of the polarization vector along the null geodesic is calculated using the Walker-Penrose integral of motion, which allows the reconstruction of the polarization vector at any point along the geodesic.  

Trajectories of photons returning to the disk are tabulated to construct the transfer functions for returning radiation ${\cal T}_{\rm ret}(r,\mu,\varphi,r_{\rm in},\mu_{\rm in}, \varphi_{\rm in},g)$, ${\cal T}_{\rm X,ret}(r,\mu,\varphi,r_{\rm in},\mu_{\rm in}, \varphi_{\rm in},g)$ and ${\cal T}_{\rm Y,ret}(r,\mu,\varphi,r_{\rm in},\mu_{\rm in}, \varphi_{\rm in},g)$, where $\mu_{\rm in}$,  $\varphi_{\rm in}$ and $r_{\rm in}$ are the incidence angles and the incidence radius. 
Convolution of these transfer functions with the disk thermal spectrum yields the spectrum and polarization of the radiation returning to the disk. 
Scattering of this radiation off the disk surface 
is calculated using the transfer matrix for reflection
defined in Section 70.3 and given in Table XXV of \citet{1960ratr.book.....C}. 
The reflected radiation is then convolved with $\cal{T}$, $\cal{T}_{\rm X}$, and $\cal{T}_{\rm Y}$ to find the observed spectrum and polarization of this reflected component. 
The \texttt{retBB} model may apply the spectral modification due to atomic processes, for which the \texttt{xillverNS} table model is used. However, note that this only modifies the spectral shape, while the polarization state remains described by Chandrasekhar's model of elastic reflection. This approach overestimates the polarization of the fluorescent and free-free component of the reflected radiation.

Figures \ref{fig:multi} and \ref{fig:isco} show example energy and polarization spectra for the parameters relevant to Cyg X-1. Note that the clockwise disk rotation assumed here is opposite to the counterclockwise rotation usually adopted in theoretical studies\footnote{We remind that in our notation $i = 180\degr - i_{\rm orb}$, where $i_{\rm orb}$ denotes the binary inclination.}. Also note that our definition of PA differs, for example, from \cite{2009ApJ...701.1175S}, who measure it from the direction parallel to the disk. 
We note an overall good qualitative agreement with previous studies of the polarization of Kerr black hole disks \citep{1980ApJ...235..224C,1990MNRAS.242..560L,2009ApJ...701.1175S}. 
At low energies, the direct disk component is dominated by emission from large radii, with the PD given by the Chandrasekhar result. 
At higher energies, the inner regions of the disk dominate, and relativistic effects reduce the PD. 
The most significant of these effects is the gravitational rotation of polarization planes, corresponding to the bending of photon trajectories. 
Combining photon beams with differently rotated polarization planes decreases PD, as can be seen from the red dotted curves in Figs.~\ref{fig:multi} and \ref{fig:isco}. 
Similar depolarization is not significant for the returning component, as it arises from a much narrower radius range. 
For this component, the bending of trajectories may affect the polarization by combining photons emitted at different $\theta$, which are differently polarized; this effect increases the PD to more than $10$\% at $E>10$~keV for the first order reflection in Fig.~\ref{fig:multi}.  
The total polarization transitions from a direct flux contribution that dominates at low energies and produces horizontal polarization, to a returning polarized flux contribution that dominates at higher energies and produces net vertical polarization.
When this transition occurs, the PD passes through a minimum, as the two contributions cancel each other out.

Figure~\ref{fig:multi} also illustrates the effect of higher-order reflection. 
For very high values of $a$, these produce a tail extending to several hundred keV with a photon index of $\Gamma \sim 3$. 
The intensity of this component is very low and, even for $a=0.998$, it contributes negligibly to the high-energy tail observed in the soft state of Cyg X-1. 
It could, however, be relevant in very soft states where only weak high-energy tails above the disk spectrum are present. 
Even in that case, its contribution to the polarized flux within the \textit{IXPE} energy band remains negligible.

As in previous studies, we considered only the reflection of radiation returning beyond the ISCO.  
Figure \ref{fig:isco} shows that reflection within the ISCO for $a=0$ is not a significant effect, and even including it, the returning component is much too weak to be measured in any real system hosting a slowly rotating black hole. 
Including the reflection within the ISCO also slightly decreases the energy at which the polarization swings from horizontal to vertical.

\begin{figure*}
\centering
\includegraphics[height=4.5cm]{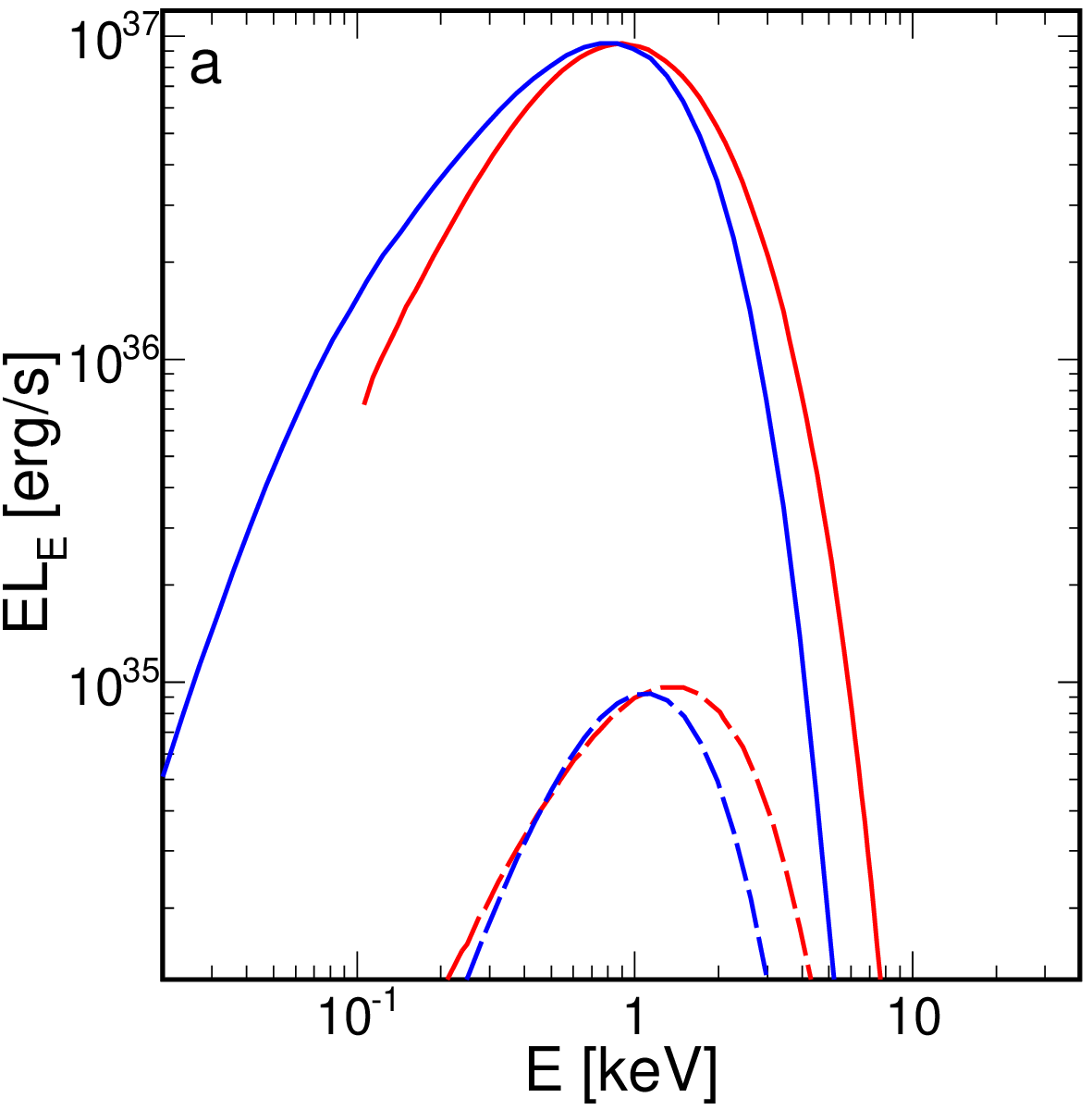} \hspace{0.1cm} \includegraphics[height=4.5cm]{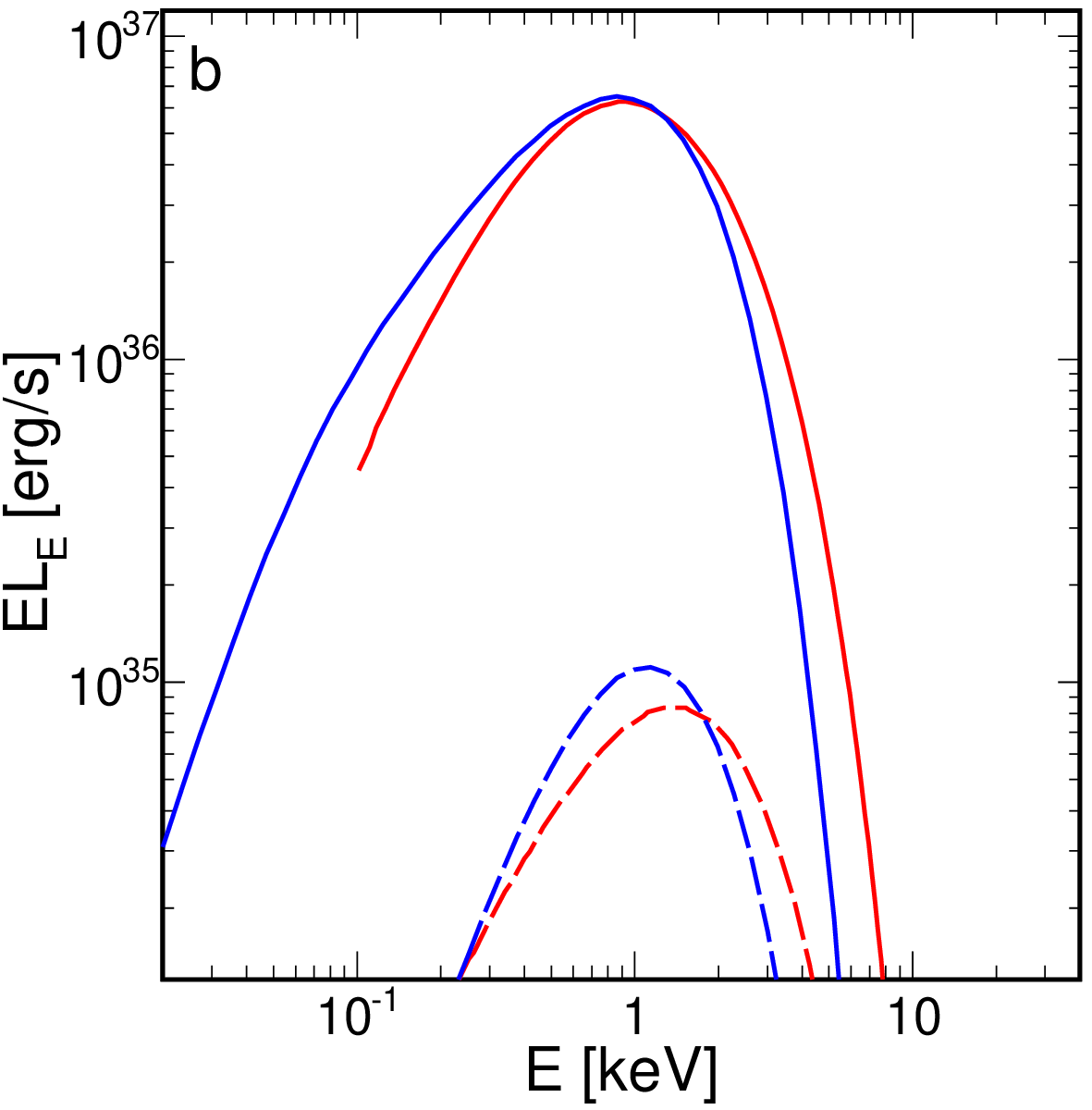}\hspace{0.1cm} \includegraphics[height=4.5cm]{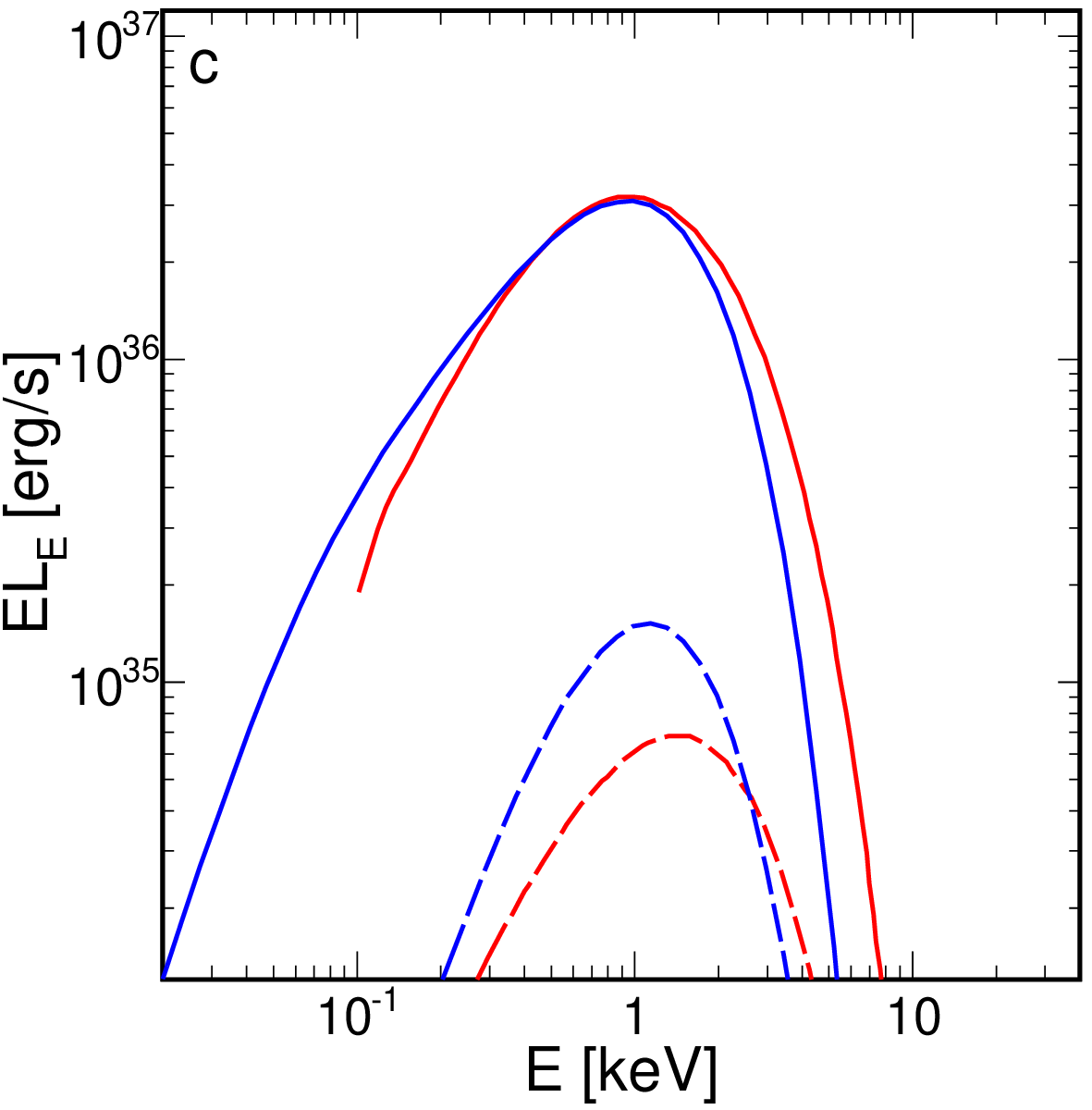}\\
\includegraphics[height=4.5cm]{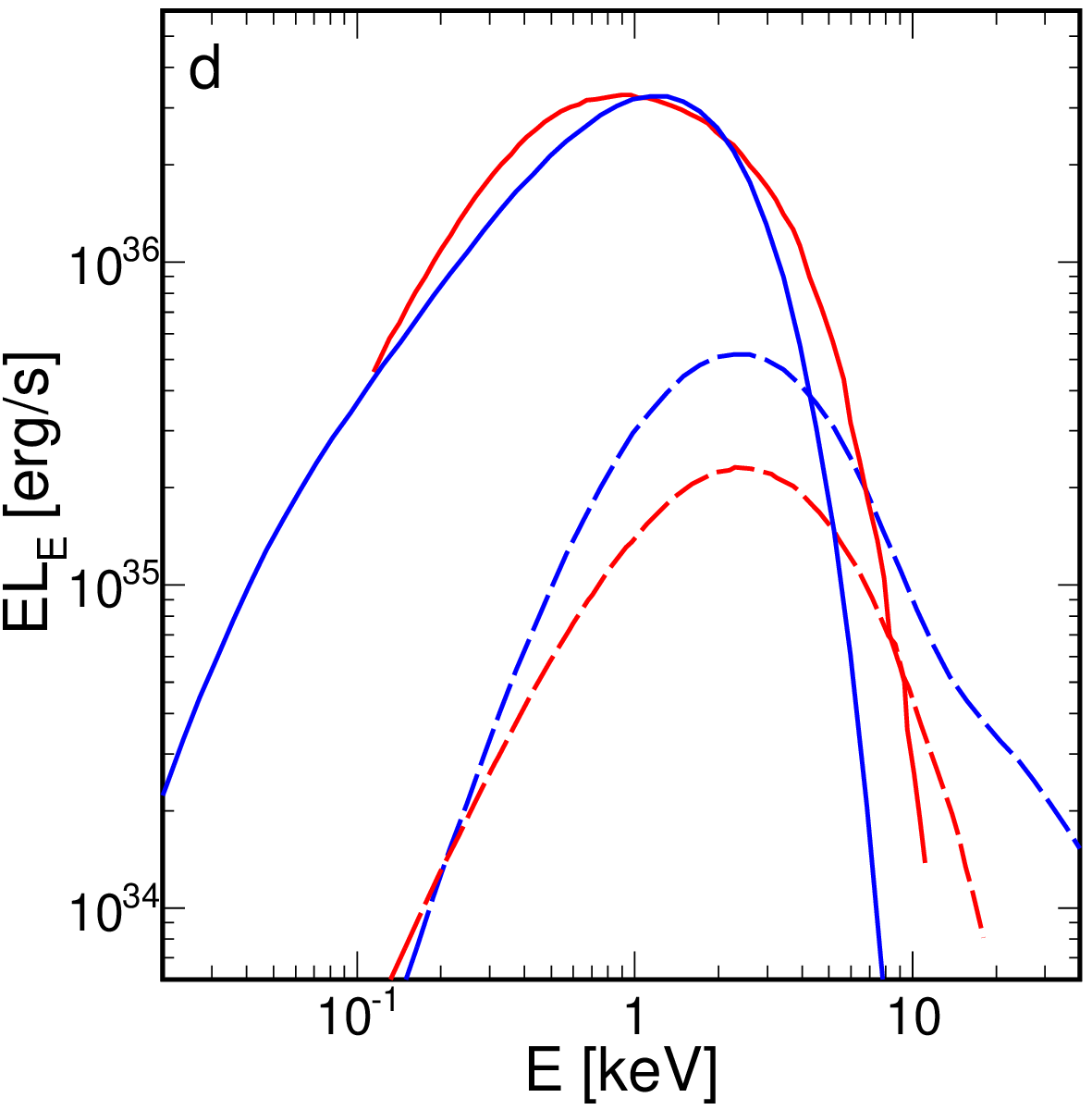} \hspace{0.1cm} \includegraphics[height=4.5cm]{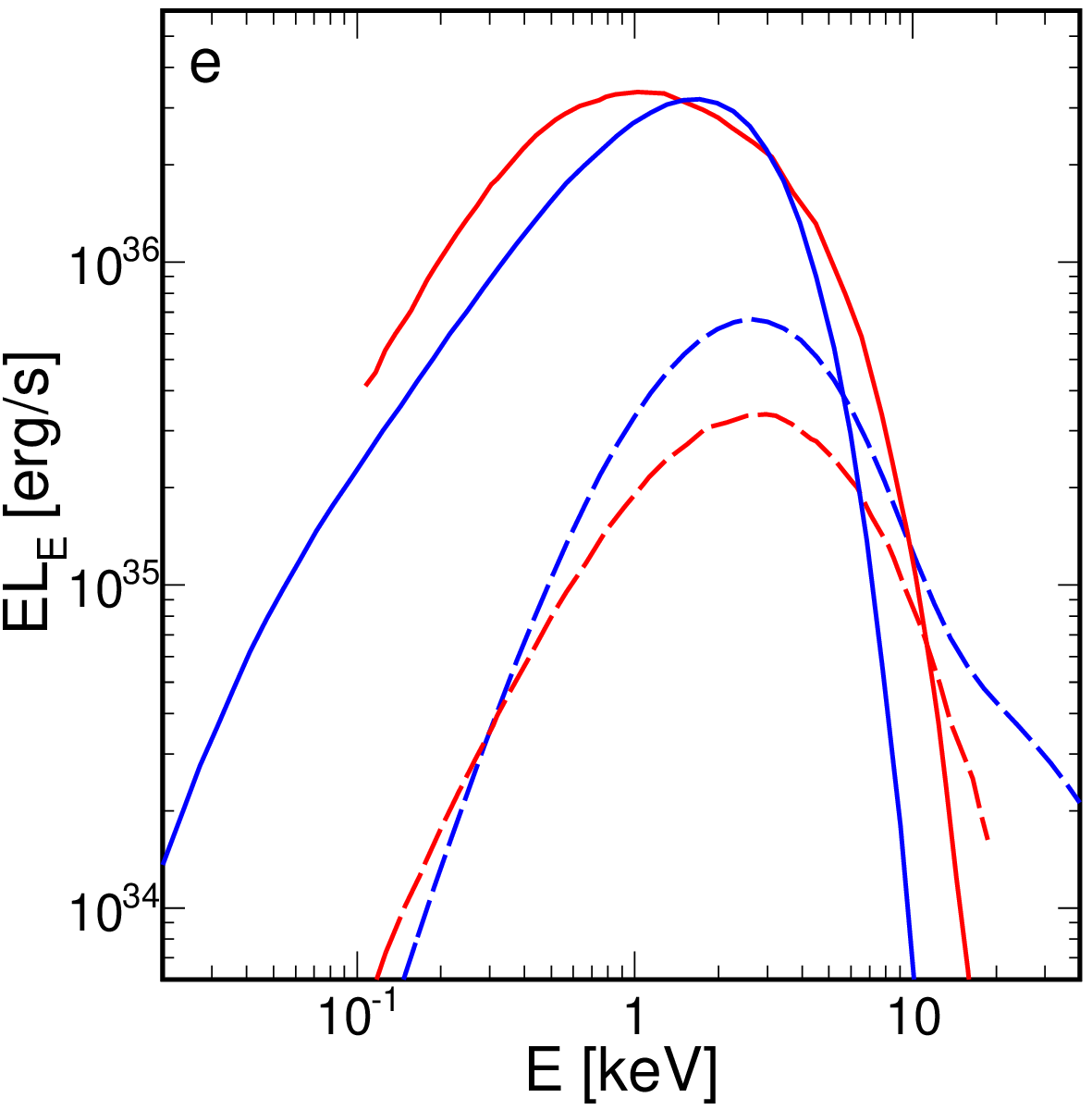}\hspace{0.1cm} \includegraphics[height=4.5cm]{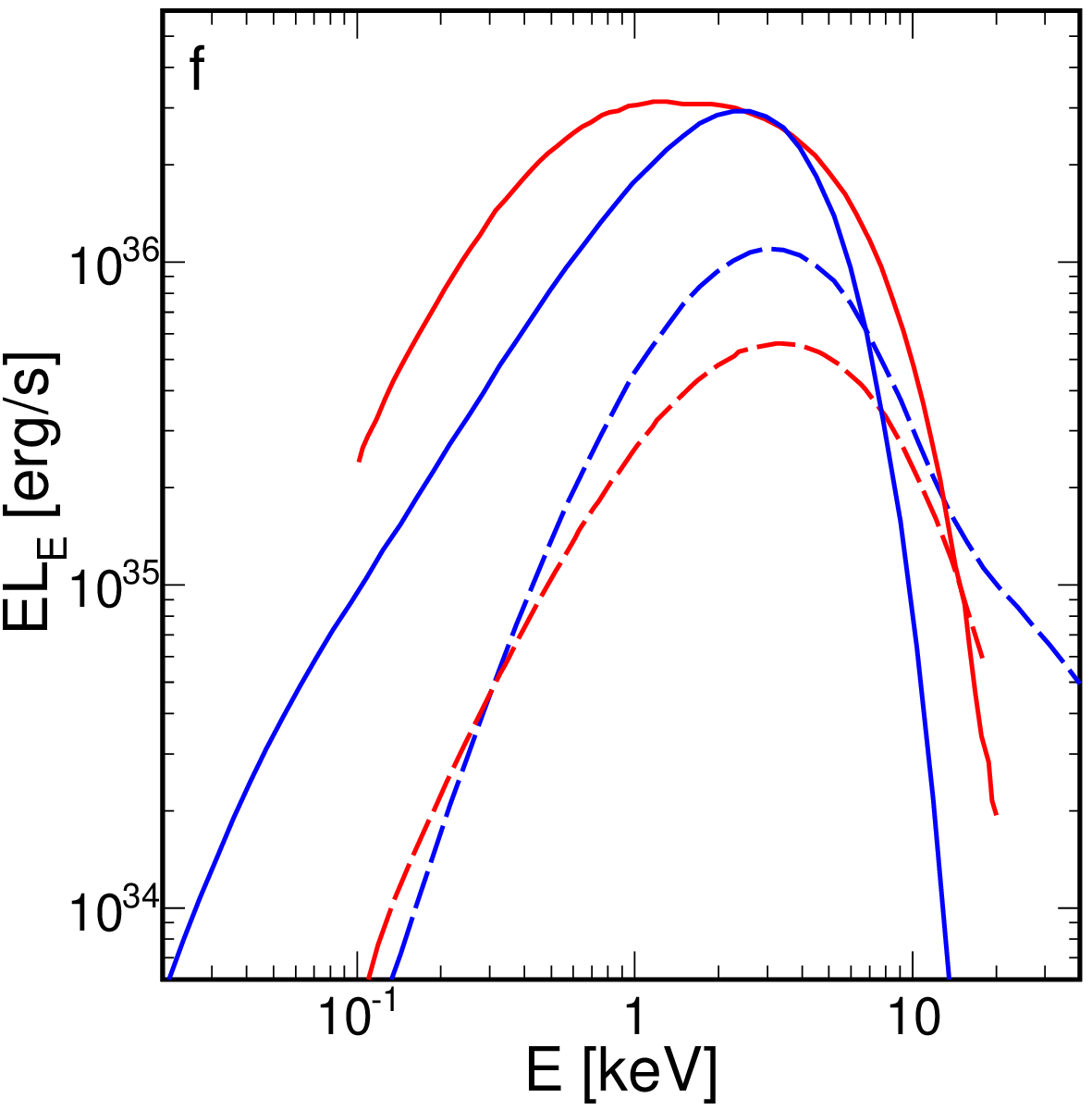}
\caption{Comparison of the spectra predicted by \texttt{retBB} (blue curves) and \cite{2009ApJ...701.1175S} 
(red curves) for $M = 10 M_\odot$, $\fcolor = 1.8$ and $i=45\degr$, $60\degr$ and $75\degr$ from left to right. In all panels, the solid curves show the direct disk radiation and the dashed curves show the elastic reflection of the returning radiation. The top panels are for $a=0$, the \cite{2009ApJ...701.1175S} 
spectra are read from the left column of their figure 4, and $\dot M = 0.2 L_{\rm Edd}/c^2$ is assumed in \texttt{retBB} to approximately match the shape thermal disk component of  \cite{2009ApJ...701.1175S}. 
 The bottom panels are for $a=0.998$, the \cite{2009ApJ...701.1175S} 
 spectra are read from the left column of their figure 6, and $\dot M = 0.02 L_{\rm Edd}/c^2$ is assumed in {\tt retBB} to approximately match the shape of the thermal disk component of \cite{2009ApJ...701.1175S}.
}
\label{fig:sk09}
\end{figure*}

We now quantitatively compare {\tt retBB} with the available predictions of other models. We first note that, for $a=0.998$, we find that $\simeq$20\% of the disk radiation returns to the disk, which agrees with previous estimates \citep[e.g.,][]{2000ApJ...528..161A,2005ApJS..157..335L}.
Figure \ref{fig:sk09} compares the energy spectra predicted by {\tt retBB} with those computed by \cite{2009ApJ...701.1175S}, digitized from their paper using WebPlotDigitizer.\footnote{https://automeris.io} Both models include multiple elastic reflections. We initially adopted the Eddington fractions reported in that work ($L = 0.1 L_{\rm Edd}$), but found discrepancies in the shapes of the direct disk spectra (while {\tt retBB} fully agrees with {\tt kerrbb}). Only a rough match could be achieved, requiring setting in {\tt retBB} very low accretion rates corresponding to disk luminosities of $< 10^{-3} L_{\rm Edd}$; these matched spectra are shown in Fig.~\ref{fig:sk09}. After normalizing the direct spectra, the luminosities of the reflected returning components agree within a factor of $\sim 2$. However, in \cite{2009ApJ...701.1175S} this component is generally weaker and, for $a=0.998$, extends to much lower energies than in {\tt retBB}.

\begin{figure}
\centering
\includegraphics[height=6.cm]{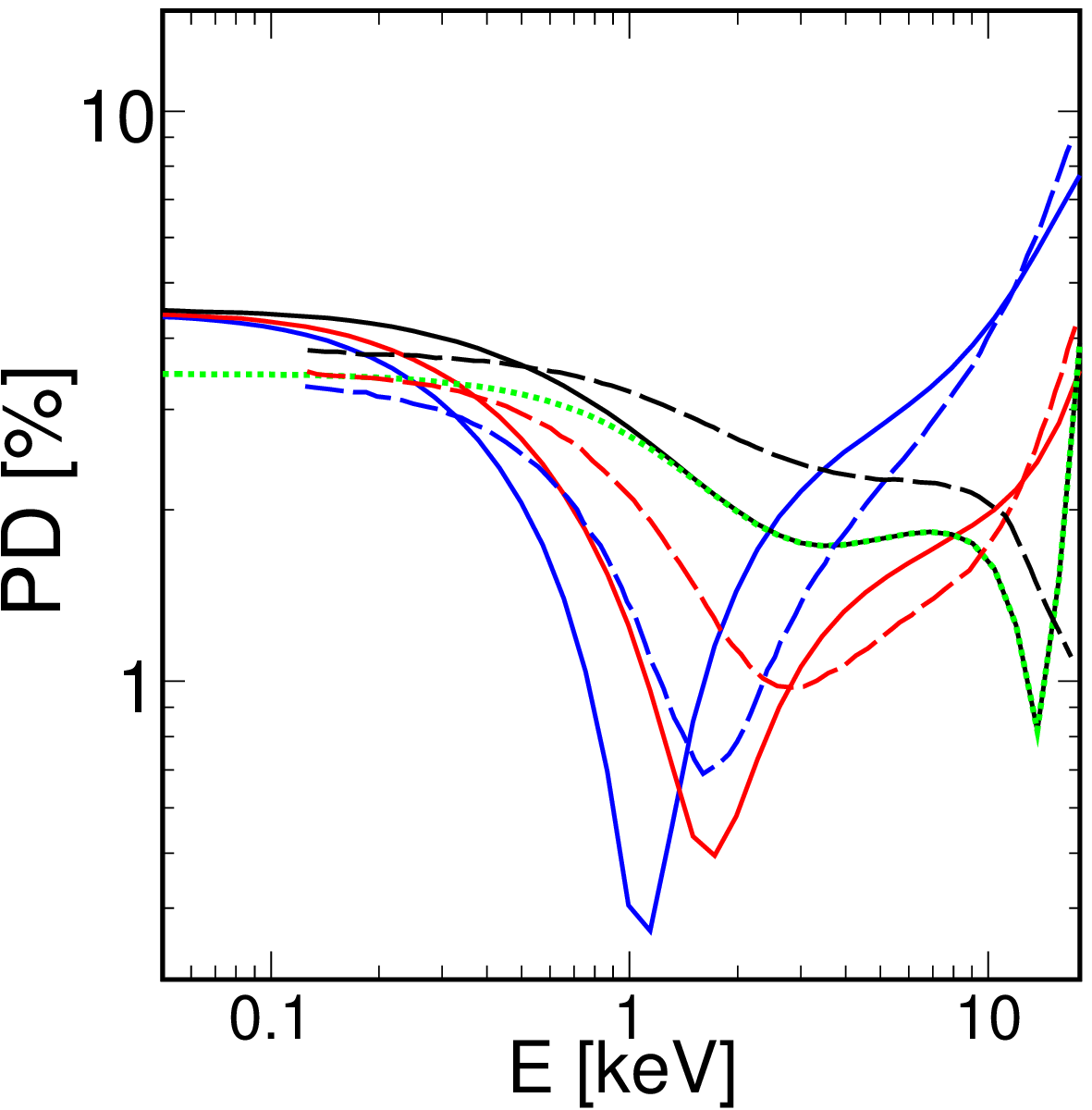}
\caption{Comparison of the polarization degrees predicted by {\tt retBB} and {\tt kerrC}. The dashed curves show PD given in figure 5 of \cite{2012ApJ...754..133K}
for $a=0$ (black), $a=0.9$ (red) and $a=0.99$ (blue). The solid curves show the PD predicted by {\tt retBB} for the same parameters ($i = 75\degr$, $\fcolor = 1.8$, $M = 10 M_\odot$, $\dot M = 2.45 \times 10^{18}$\,g\,s$^{-1}$ for $a=0$, $\dot M=9 \times 10^{17}$\,g\,s$^{-1}$ for $a=0.9$ and $\dot M=5.3 \times 10^{17}$\,g\,s$^{-1}$ for $a=0.99$) for a single elastic reflection (using a multiple reflection does not improve the agreement). The green dotted curve shows the {\tt retBB} spectrum with $R_{\rm out} = 100$. 
}
\label{fig:kerrc}
\end{figure}

We now compare {\tt retBB} with {\tt kerrC} of \cite{2012ApJ...754..133K}. The published {\tt kerrC} results do not separate the direct and reflected components, preventing a comparison of energy spectra like those in Fig.~\ref{fig:sk09}. 
Figure~\ref{fig:kerrc} compares the PDs predicted by the two models. 
We find moderate differences between {\tt retBB} and {\tt kerrC}, with the latter extracted from the PD-plots of \cite{2012ApJ...754..133K} using WebPlotDigitizer. 
At low energies, the Chandrasekhar limit, ${\rm PD} \simeq 4.5\%$ for $i = 75\degr$, should be recovered. 
This is satisfied by {\tt retBB}, whereas {\tt kerrC} slightly underpredicts PD, likely due to assuming a relatively small outer radius ($R_{\rm out} = 100$). 
This effect is illustrated by the green dotted {\tt retBB} curve. 
For the parameters considered here, differences related to $R_{\rm out}$ affect only energies below the \textit{IXPE} band.

\end{appendix}
\end{document}